\documentclass{svjour3}                     % onecolumn (standard format)
\smartqed  % flush right qed marks, e.g. at end of proof
\usepackage{graphicx}
\usepackage{mathptmx}      % use Times fonts if available on your TeX system
\usepackage{geometry}
\usepackage{hyperref}
\usepackage{relsize}

\usepackage{bigints}            

\usepackage{algorithm,algorithmic}
\usepackage{amssymb}
\usepackage{amsmath}
\usepackage{amsfonts}
\usepackage{fullpage}
\usepackage{latexsym}
\usepackage{setspace}
\usepackage{float}
\usepackage{graphicx}
\usepackage{caption}
\usepackage{subcaption}
\usepackage{multirow}
\usepackage{array}
\usepackage{mathtools}
\usepackage{url}

\usepackage{enumerate}

\usepackage{booktabs}

\usepackage{tikz} \usetikzlibrary {plotmarks}

\usepackage{paralist}
\usepackage{array}
\usepackage[mathscr]{euscript}

\usepackage{mathtools}

\usepackage{bm}
\newcommand{\hf}{\frac12}							% 1/2

\newcommand{\muinv}{\mu^{-1}}						% mu inverse
\newcommand {\R}   {\mathbb{R}}   					% real space
\newcommand {\Hdiv} {\mathcal{H}(div;\Omega)}		% H-div
\newcommand {\Hcurl}{\mathcal{H}(curl;\Omega)}		% H-curl
\newcommand{\B}{{\vec{B}}}							% magnetic flux
\newcommand{\E}{\vec{E}}							% electric field
\newcommand{\F}{\vec{F}}							% face function - in Hdiv
\newcommand{\J}{\vec{J}}							% electric current density
\newcommand{\Js}{\J_s}								% source current density
\newcommand{\W}{{\vec{W}}}							% edge function - in Hcurl
\newcommand{\curl}{\ensuremath{ \nabla \times\,}}	% curl operator
\newcommand{\tangentv}{\pmb{\tau}}					% unitary tangential vector
\newcommand{\normalv}{\vec{n}}						% unitary normal vector
\newcommand{\coarseM}{\mathscr{M}^H}						% coarse mesh
\newcommand{\fineM}{\mathscr{M}^h}						% fine mesh

\newcommand{\OmegakH}{\Omega_k^H}					% kth coarse mesh cell 
\newcommand{\OmegakHext}{\Omega_k^{H,{\sf ext}}}	    % kth extended coarse mesh cell 

\newcommand{\bfb}{{\bf b}}						% discrete magnetic flux
\newcommand{\bfe}{{\bf e}}						% discrete electric field
\newcommand{\bfq}{\bf q}						% auxiliary source term
\newcommand{\bfSig}{\pmb \Sigma}				% grid function of Sigma
\newcommand{\bfmu}{\pmb \mu}					% grid function of mu
\newcommand{\bfmuinv}{\pmb \muinv}				% grid function of mu inverse
\newcommand{\CURL}{\bf CURL}					% discrete curl operator
\newcommand{\Mmuinv}{{\bf{M_f}}(\bfmuinv)} 		% face mass matrix of mu inverse
\newcommand{\MSig}{{\bf{M_e}}(\bfSig)} 			% edge mass matrix of sigma

\newcommand{\bfC}{{\bf C}}						% Mesh connectivity
\newcommand{\bfI}{{\bf I}}						% Identity matrix
						
\newcommand{\bfA}{{\bf A}}						% Matrix of Maxwell's system 
\newcommand{\bfASig}{{\bfA(\bfSig)}}			% Matrix of Maxwell's system with dependency of sigma
\newcommand{\bPhi}{{\pmb \Phi}}

\newcommand{\PHh}{{\bf P}}						% global coarse-2-fine interpolation matrix
\newcommand{\Pk}{{{\bf P}_k}}					% coarse-2-fine interpolation matrix for kth coarse cell
\newcommand{\mA}{\hat{\mathcal{A}}}

\journalname{No-name}

\begin{document}

\title{An oversampling technique for the multiscale finite volume method to simulate electromagnetic responses in the frequency domain}

\titlerunning{Multiscale finite volume with oversampling for Maxwell's equations}        

\author{Luz Ang\'elica Caudillo-Mata \and Eldad Haber \and Christoph Schwarzbach}

\institute{   L. A. Caudillo-Mata\at
              \email{lcaudill@eos.ubc.ca} \\ \\
              E. Haber \at
              \email{ehaber@eos.ubc.ca} \\ \\
              C. Schwarzbach \at
              \email{cschwarz@eos.ubc.ca} \\ \\
              Geophysical Inversion Facility \at 
              Earth, Ocean and Atmospheric Sciences Department \at
              University of British Columbia \at
              4013-2207 Main Mall, Vancouver, BC, Canada, V6T 1Z4
          }

\date{Received: date / Accepted: date}

\maketitle

\begin{abstract}
In order to reduce the computational cost of the simulation of electromagnetic responses in geophysical settings that  involve highly heterogeneous media, we develop a multiscale finite volume method with oversampling for the quasi-static Maxwell's equations in the frequency domain.  We assume a coarse mesh nested within a fine mesh that accurately discretizes the problem. For each coarse cell, we independently solve  a local version of the original Maxwell's system subject to linear boundary conditions on an extended domain, which includes the coarse cell and a neighborhood of fine cells around it.  The local Maxwell's system is solved using the fine mesh contained in the extended domain and the mimetic finite volume method.  Next, these local solutions (basis functions) together with a weak-continuity condition are used to construct a coarse-mesh version of the global problem.  The basis functions can be used to obtain the fine-mesh details from the solution of the coarse-mesh problem.
 Our approach leads to a significant reduction in the size of the final system of equations and the computational time, while accurately approximating the behavior of the fine-mesh solutions.  We demonstrate the performance of our method using a synthetic 3D example of a mineral deposit.
\end{abstract}

%------------------------------------------------------------
% SECTION
%------------------------------------------------------------
\section{Introduction} \label{intro}

% Motivation and problem statement
Accurate and efficient simulation of electromagnetic (EM) responses --- EM fields and fluxes --- in large-scale heterogeneous media is crucial to the exploration and imaging of geological formations in a wide range of geophysical applications, including mineral and hydrocarbon exploration, water resource utilizations, and geothermal pow\-er extractions (cf. \cite{Oldenburg1990,Zhdanov2010}).  
One major challenge in practice to perform this type of simulation is the excessive computational cost it involves.
Realistic geophysical settings often consider large computational domains, features that vary at multiple spatial scales, and a wide variation over several orders of magnitude of the geological properties of the media. 
Since all of these factors can have a significant impact on the behavior of the EM responses of interest, if we wish to obtain an accurate approximation to the responses, the mesh used in classical discretization techniques, such as finite volume (FV) or finite element (FE), must capture the structure of the heterogeneity present in the setting with sufficient detail.  This leads to the use of very large meshes that translate into solving huge systems of equations ---  in some cases,  in the order of billions of unknowns.     

% Alternatives: adaptive mesh refinement approaches (overview,gaps)
Adaptive mesh refinement approaches have been used to overcome the computational cost of realistic EM simulations (cf. \cite{Lipnikov2004,Haber2007a,Schwarzbach2009,Horesh2011,Key2011a}).
Although these ap\-proach\-es have produced accurate approximations to the EM responses at an affordable cost, they face one major issue: the mesh must still capture the spatial distribution of the media heterogeneity both inside and outside the region where we measure the EM responses. This restricts the ability of these approaches to reduce the size of the system to be solved.

% On the other hand: MS FV/FE as an alternative to reduce size of the system, hence reduce computational cost - for elliptic problems
Alternatively, multiscale FV/FE techniques aim to reduce the size of the linear system by constructing a coarse-mesh version of the fine-mesh system that is much cheap\-er to solve. These techniques can be classified within the family of Model Order Reduction methods, where the resulting fine-mesh system from the discretization of the partial differential equation (PDE) is replaced by its projected form (cf. \cite{Benner2015}). 
Multiscale FV/FE techniques have been extensively studied in the field of modeling flow in heterogeneous porous media, where they have been successfully used to drastically reduce the size of the linear system while producing accurate solutions similar to that obtained with FE or FV discretization schemes on a fine mesh (cf. \cite{Efendiev2007,Pavliotis2008}). Researchers in this field have noted that the projection matrix constructed using multiscale FV/FE methods may lead to numerical solutions that contain `resonance errors', that is errors that appear when the coarse-mesh size and the wavelength of the small scale oscillation of the media heterogeneity are similar (cf. \cite{Hou1997,Hou1999,Efendiev2009}).  A solution in such case is to use oversampling techniques in the construction of the projection matrix.  Haber and Ruthotto \cite{Haber2014a} extended multiscale FV techniques for application in EM modeling.  However, their work did not include oversampling and therefore, their technique is affected by the same resonance problems obtained when using the multiscale technique for the flow problem.

% 6) Contribution of this paper - oversampling technique
% Make clear the scope and novelty of this work 
% Connect to how we solve the gaps with this technique: for adaptive mesh refinement we now can coarse the mesh outside the survey area, for upscaling - say MSFV is a more systematic approach than upscaling
% Multiscale methods are applicable to general multiple-scale problems without restrictive assumptions.  
Recognizing the success of oversampling techniques in fluid flow applications, in this paper, we extend their use for application in EM modeling, where being able to reduce the size of the problem can be particularly advantageous when large domains are considered or when the mesh must capture the spatial distribution of the media heterogeneity outside the region where the EM responses are measured.  In particular, we propose an oversampling technique for the multiscale FV method introduced by \cite{Haber2014a} for the quasi-static Maxwell's equations in the frequency domain.  
% Summarize at high level how oversampling works - octree meshes
We show that our method produces more accurate solutions than the multiscale FV method without oversampling. % Say that this method is new and there are not much research out there targeting EM modeling

% 6) Organization
This paper is organized as follows. Section \ref{sec:background} introduces the mathematical model used and provides an o\-ver\-view of the mimetic finite volume discretization meth\-od, which is used as a building block to develop our oversampling technique.  Section \ref{sec:msfvo} presents the development of the oversampling technique proposed for the multiscale finite volume method introduced by \cite{Haber2014a}.  Section \ref{sec:numres} demonstrates the performance of our oversampling technique using a synthetic 3D example of a mineral deposit.  Finally, Section \ref{sec:conclusions} concludes the paper by discussing the capabilities and limitations of our method. 

%------------------------------------------------------------
% SECTION
%------------------------------------------------------------
\section{Mathematical background} \label{sec:background}

% Overview section
This section introduces the mathematical model we focus on in this work and provides an overview of the mimetic finite volume discretization method, which is used as a building block to develop the oversampling technique we propose.

%------------------------------------------------------------
\subsection{The quasi-static Maxwell's equations in the frequency domain} \label{sec:mathModel}

% Introduce Maxwell's system used
We focus on EM geophysical problems in the frequency domain where the quasi-static approximation applies, that is, for the frequency range we work on, the electrical current displacement can be safely neglected when compared with the electrical current density (cf. \cite{Ward1988}). For this scenario, the governing mathematical model is given by the first-order form of the Maxwell's equations: 
\begin{eqnarray}
\curl \E + \imath \omega \B       &= \ \vec{0}, \ & \mbox{in} \ \ \Omega, \label{eq:me1} \\
\curl \mu^{-1}\B - \Sigma \E &= \ \Js, \ & \mbox{in} \ \ \Omega, \label{eq:me2}
\end{eqnarray}
where $\E$ denotes the electric field, $\B$ denotes the magnetic flux density, $\Js$ denotes the source term, $\omega$ denotes the angular frequency, $\imath$ is the unit imaginary number, and $\Omega \subset \R^3$ denotes the domain. The PDE coefficients, $\mu$ and $\Sigma$, are the magnetic permeability and electrical conductivity, respectively.   We assume that both coefficients are $3 \times 3$ symmetric positive definite (SPD) tensors that vary over multiple spatial scales and several orders of magnitude. In particular, this assumption implies that the PDE coefficients are highly discontinuous in the domain.
These PDE coefficients model the anisotropic and highly heterogeneous behavior of the medium in the geophysical problem we consider. We refer to them as the {\em medium parameters}.

% Introduce boundary conditions used
The Maxwell's system \eqref{eq:me1}-\eqref{eq:me2} is typically closed with natural boundary conditions of the form
\begin{equation} \label{eq:naturalBC}
 \mu^{-1} \B \times \normalv = \vec{0} \ \ \mbox{or} \ \ (\curl \E) \times \normalv = \vec{0}, \ \ \mbox{on} \ \partial \Omega,   
\end{equation}
or with non-homogeneous Dirichlet boundary conditions given by 
\begin{equation} \label{eq:dirichletBC}
    \E \times \normalv =  \E_0 \times \normalv, \ \mbox{on} \ \partial \Omega, 
\end{equation}
were $\partial \Omega$ denotes the boundary of $\Omega$, $\normalv$ denotes the unit outward-pointing normal vector to $\partial \Omega$, and $\E_0$ specifies the tangential components of $\E$ at $\partial \Omega$. However, more general boundary conditions can be imposed to the Maxwell's system as discussed in \cite{Jin2002,Ward1988}.

%------------------------------------------------------------
\subsection{Mimetic finite volume method} \label{sec:mfv}

% MFV overview: definition, advantages  
Since the mimetic finite volume (MFV) method is a building block to develop the oversampling technique we propose in this work, an overview of this method is provided in this section. 
Full derivation details can be found in \cite{Hyman1998,Hyman1999,Hyman1999a,Haber2014}.  
MFV is an extension of Yee's method (\cite{Yee1966}) that constructs discrete curl, divergence and gradient operators satisfying discrete analogs of the main theorems of vector calculus  involving such operators. Therefore, the discrete differential operators obtained with MFV do not have spurious solutions and the ``divergence-free'' magnetic field condition for Maxwell's equations is automatically satisfied. 
In addition, MFV assumes that the tangential components of $\E$ and the normal components of $\B$ are continuous trough media interfaces. This choice guarantees that problems with strongly discontinuous PDE coefficients are treated properly. Furthermore, MFV leads to sparse and symmetric linear systems of equations.

% Introduce weak-form and its elements 
The MFV method begins by considering the weak form of the system (\ref{eq:me1})-(\ref{eq:me2}), given by
\begin{eqnarray}
(\curl \E, \F) + \imath \omega (\B,\F) &=& 0,  \label{eq:wem1} \\
(\curl \mu^{-1}\B,  \W)- (\Sigma \E,\W) &=& (\Js,\W),  \label{eq:wem2}
\end{eqnarray}
where $\F \in \Hdiv$ and $\W \in \Hcurl$ are test functions; $\Hdiv$ and $\Hcurl$ are the Hilbert spaces of square-integrable vector functions on $\Omega$ with square-integrable divergence and curl, respectively, and $(\cdot, \cdot)$ denotes the inner product given by $(\vec{P},\vec{Q}) = \int_{\Omega} P^x\overline{Q^x} + P^y\overline{Q^y} \\ + P^z\overline{Q^z} \, dV$. 

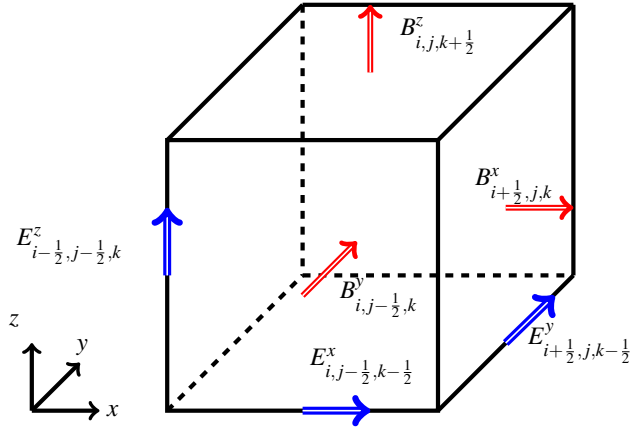
\begin{figure}
\begin{center}
\begin{tikzpicture}[thick, scale=0.90, every node/.style={transform shape}]
  % --- Draw axis

  % Lines
  \draw[ultra thick,->]  (-2,0) -- (-1,0);  %x axis
  \draw[ultra thick,->]  (-2,0) -- (-1.29,0.7071);  %y axis
  \draw[ultra thick,->]  (-2,0) -- (-2,1);  %z axis

  %Labels for axis
  \pgfputat{ \pgfxy(-0.8,0)}{\pgfbox[left,center]{$x$} };
  \pgfputat{ \pgfxy(-2.1,1.2)}{\pgfbox[left,center]{$z$} };
  \pgfputat{ \pgfxy(-1.2,0.85)}{\pgfbox[left,center]{$y$} };

  % --- Draw labeled control volume

  \draw[ultra thick]  (0,0) -- (4,0) -- (4,4) -- (0,4) -- cycle;  % whole front face
  \draw[dashed, ultra thick]  (0,0) -- (2,2);

  \draw[dashed, ultra thick] (2,2) -- (6,2);
  \draw[dashed, ultra thick] (2,6) -- (2,2);

  \draw[ultra thick]  (6,6) -- (2,6);
  \draw[ultra thick]  (6,2) -- (6,6);
  \draw[ultra thick]  (4,0) -- (6,2);                             % edge bottom face, right side
  \draw[ultra thick]  (0,4) -- (2,6);                             % edge upper face, left side
  \draw[ultra thick]  (4,4) -- (6,6);                             % edge upper face, right side

        \draw[red, thick, double,->] (5,3) -- (6,3);
        \draw[red, thick, double,->] (3,5) -- (3,6);
        \draw[red, thick, double,->] (2.0,1.7) -- (2.8,2.5);  %By

        \draw[blue, very thick, double,->] (0,2) -- (0,3);
        \draw[blue, very thick, double,->] (2,0) -- (3,0);
        \draw[blue, very thick, double,->] (5,1) -- (5.72,1.7);    %Ez

     \pgfputat{\pgfxy(4.1,3.0)}{\pgfbox[left, center]{$B^x_{i+\hf,j,k}$}};
   \pgfputat{\pgfxy(3.1,5)}{\pgfbox[left, center]{$B^z_{i,j,k+\hf}$}};
   \pgfputat{\pgfxy(2.3,1.5)}{\pgfbox[left, center]{$B^y_{i,j-{\frac 12},k}$}};
   \pgfputat{\pgfxy(1.9,0.6)}{\pgfbox[left,center]{$E^x_{i,j-\hf,k-\hf}$}};
   \pgfputat{\pgfxy(-2.0,2.2)}{\pgfbox[left,center]{$E^z_{i-\hf,j-\hf,k}$}};
     \pgfputat{\pgfxy(4.8,0.9)}{\pgfbox[left,center]{$E^y_{i+\hf,j,k-\hf}$}};

\end{tikzpicture}
\caption{Control volume cell showing the staggered discretization for $\E$ on its edges, $\B$ on its faces, and the medium parameters $\mu$ and $\Sigma$ at the cell-centers. \label{fig:YeeCell}} 
\end{center}
\end{figure}

% Staggered mesh, impose natural boundary conditions, integration of inner products using quadrature rules, how to compute b
Next, the method continues by using a staggered mesh to discretize $\E$ on the edges, $\B$ on the faces, and the PDE coefficients $\mu$ and $\Sigma$ at the cell-centers. Figure \ref{fig:YeeCell} shows a control volume cell with the allocation of these variables.  When the natural boundary conditions \eqref{eq:naturalBC} are imposed to the system \eqref{eq:wem1}-\eqref{eq:wem2}, and their corresponding inner products are computed using low order quadrature formulas, the MFV method yields the following linear system
\begin{eqnarray} \label{eq:mfvSystemNatBC} 
\bfASig \bfe = \left(\CURL^{\top} \Mmuinv \CURL + \imath \omega \MSig\right) \bfe \nonumber \\  = -\imath \omega \bfq,
\end{eqnarray}
where $\bfSig$, $\bfmu$ and $\bfe$ are the discrete approximations at the corresponding mesh points for $\Sigma$, $\mu$ and $\E$, respectively; $\bfq$ is the resulting discretization for the source term $\Js$. Additionally, $\CURL$, $\Mmuinv$ and $\MSig$ are the corresponding discrete operators for the continuous operator $\curl$ and the mass matrices for the medium parameters $\mu$ and $\Sigma$, respectively. 

% System of equations with natural bc: complex, sparse, symmetric and ill-conditioned when omega or sigma are small
% Point towards review on what solvers to use.
The matrix $\bfASig$ of the system \eqref{eq:mfvSystemNatBC} is complex, sparse, symmetric, and, in practice, it tends to be severely ill-conditioned for the cases where the geophysical setup includes very low conductivity values (e.g. when air is considered in the setup) or the survey considers very low frequencies.  For a review on direct and iterative solvers that can be used to solve this system see \cite{Haber2014,Haber2001,Hiptmair1998}.

% System of equations with non-homogeneous Dirichlet boundary conditions
To impose the non-homogeneous Dirichlet boundary conditions \eqref{eq:dirichletBC}, which imply the values of the tangential components of the electric field at the boundary are known, the matrix $\bfASig$ and the vectors $\bfe$ and $\bfq$ from equation \eqref{eq:mfvSystemNatBC} are reordered into interior  edges ($\sf ie$) and boundary edges ($\sf be$).  Thus, the system to be solved in terms of the unknown $\bfe^{\sf ie}$ is
\begin{equation}
\bfA^{\sf ie,\sf ie}\bfe^{\sf ie} = - \left(\imath \omega \bfq^{\sf ie} +\bfA^{\sf ie,\sf be}\bfe^{\sf be}\right), \label{eq:system}
\end{equation}
where $\bfA^{\sf ie,\sf ie}$, $\bfA^{\sf ie,\sf be}$, and ${\bf q^{\sf ie}}$ represent the corresponding partitions of the matrix $\bfASig$ and the vector $\bf q$ of the system \eqref{eq:mfvSystemNatBC}, $\bfe^{\sf ie}$ is the discretized electric field at the interior edges, and $\bfe^{\sf be}$ is the discretized electric fields at the boundary. 

% Say how to compute b
Once we compute $\bf e$, we can compute the discrete magnetic flux at the mesh faces, $\bfb$, using the discrete version of equation \eqref{eq:me1}, as follows
\begin{equation}
   \bfb =  -\frac {1}{\imath \omega} \, \CURL \, \bfe. \label{eq:b} \\
\end{equation}

% TODO:  Put labels for xi, xi+1, the same for y and z
% Figure : Yee-cell with E-B discretization
%----------------------------------------------------------------------

%------------------------------------------------------------
% SECTION
%------------------------------------------------------------
\section{Multiscale finite volume method with oversampling} \label{sec:msfvo}

This section provides an overview of the multiscale finite volume method for the quasi-static Maxwell's equations in the frequency domain, discusses the need of an oversampling technique to increase the accuracy of the solution obtained with such method, and introduces the oversampling technique we propose. 

%------------------------------------------------------------
\subsection{Multiscale finite volume method for EM modeling} \label{sec:msfv}

% Brief description of multiscale methods 
Multiscale FV/FE methods have been extensively used to improve the computational performance of simulating physical problems in petroleum engineering and composite materials that contain multiple spatial scales and whose physical properties vary over several orders of magnitude. 
These multiscale methods can reduce drastically the size of the fine-mesh system from the discretization of the PDE by constructing a coarse-mesh version of the system that is much cheaper to solve.  The solutions obtained with multiscale methods capture effectively the small scale effect on the large scales, without having to resolve all the small scale features present in the problem.  In addition, the multiscale solutions achieve a level of accuracy similar to that obtained with traditional discretization schemes (e.g. FV or FE) on a fine mesh. At present, developing efficient multiscale methods using different discretization schemes and tailoring them for use to diverse applications is an active research area (cf. \cite{Efendiev2009,Hajibeygi2009,Hou2003,Hou1997,Jenny2003,MacLachlan2012,Pavliotis2008}).

% Multiscale FV for EM modeling
Haber and Ruthotto (\cite{Haber2014a}) adapted the general lines proposed by Hou and Wu (\cite{Hou1997}), Jenny et al., (\cite{Jenny2003}), and MacLachlan and Moulton (\cite{MacLachlan2006}), where multiscale FE and FV methods are developed for elliptic problems with strongly discontinuous coefficients, to develop a multiscale finite volume (MSFV) method that fits the staggered discretization of vector fields typically used in the MFV discretization method. Since we use the MSFV meth\-od as a building block for the oversampling technique we propose in this work, we provide next an overview of this method, which can be summarized in the following four steps.

% FIGURE
\begin{figure*}[htb!]
  \centering \includegraphics[width=1.0\textwidth]{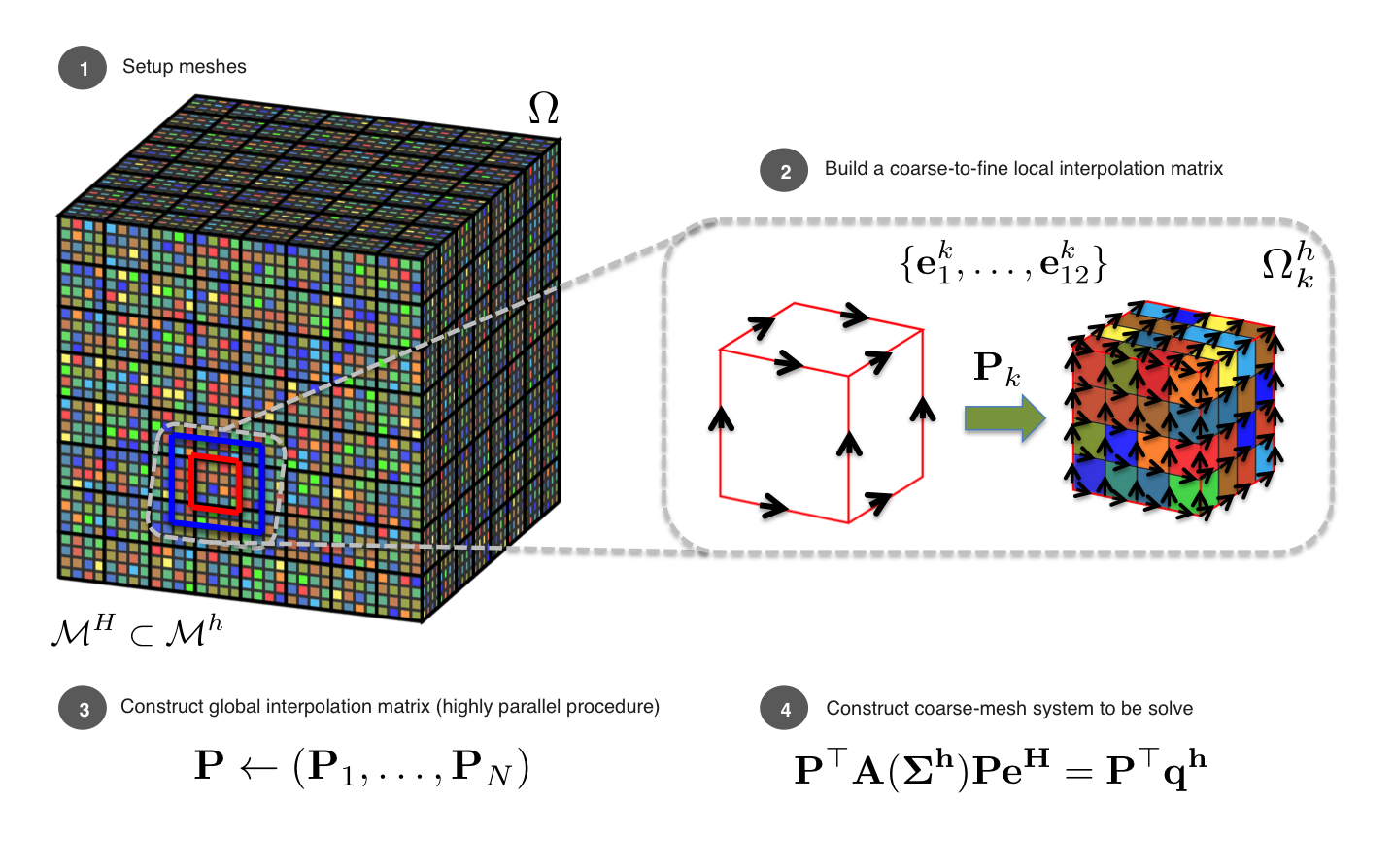}
  \caption{Schematic representation of the procedure to implement the MSFV method. \label{fig:msfv}}
\end{figure*}

% Step 1:  mesh setup and notation
First, let us assume a coarse mesh, $\coarseM$, nested into a fine mesh, $\fineM$, i.e., $\coarseM \subseteq \fineM$ (Step 1 in Figure \ref{fig:msfv}). Where $\fineM$ accurately discretizes the features in the model where the electrical conductivity varies, and $\coarseM$ is a user-chosen mesh that typically is much coarser than the fine mesh and satisfies the guidelines for mesh design in the areas where the EM responses are measured. In particular, $\coarseM = \cup_{k=1}^N \OmegakH$, where $N$ is the number of coarse-mesh cells and $\OmegakH$ denotes the kth coarse-mesh cell; $\fineM = \cup_{i=1}^n \Omega_i^h$, where $n$ is the number of fine-mesh cells and $\Omega_i^h$ denotes the ith fine-mesh cell; and $N \ll n$. The MSFV method was originally developed for nested tensor meshes, we will show an example in Section \ref{sec:numres} where we use nested OcTree meshes as the mesh setup. 

\begin{figure}[htb!]
  \centering \includegraphics[width=0.8\textwidth]{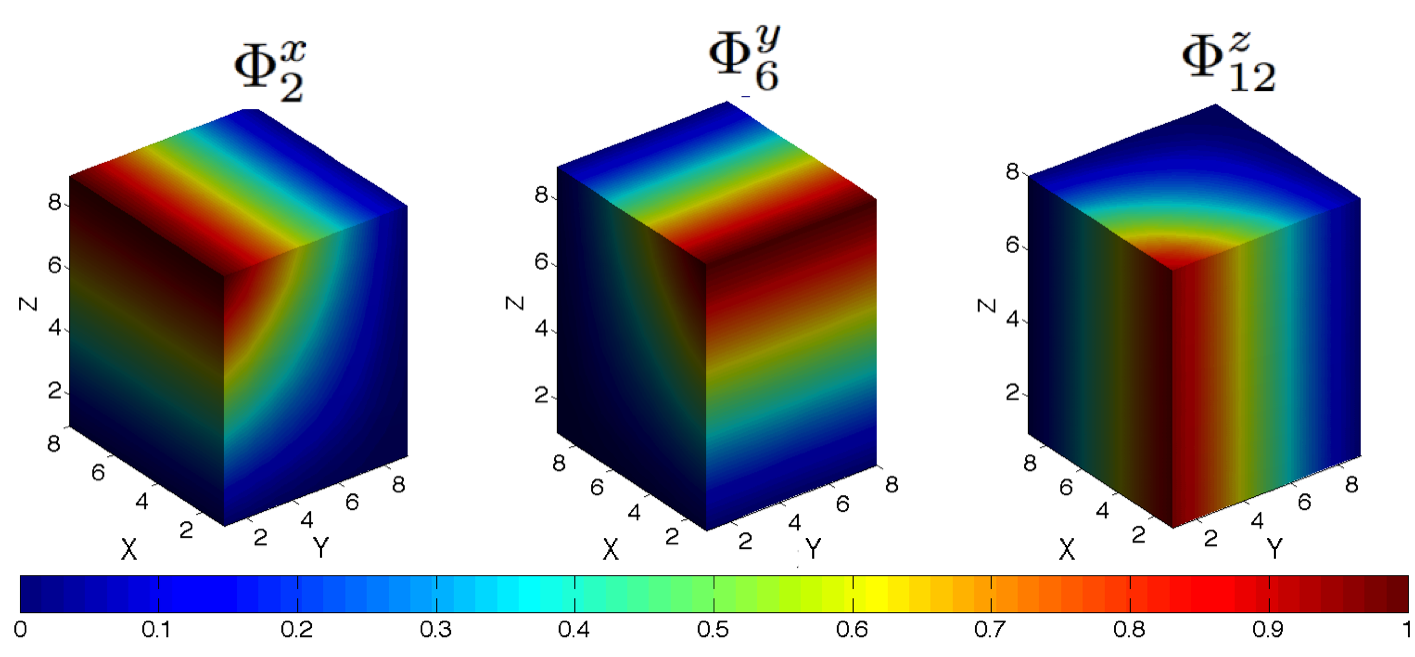}
  \caption{Non-zero components for three out of the twelve linear boundary conditions $\{\bPhi_l\}_{l=1}^{12}$. Each $\bPhi_l$ is a vector field that takes the value 1 along the tangential direction to the lth edge of $\OmegakH$ and decays linearly to 0 in the normal directions to the same edge. \label{fig:linearBC}}
\end{figure}

% Step 2: Solve twelve local problems to compute local interpolation matrices
Second, for each coarse-mesh cell, $\OmegakH; \ k=1,\ldots,N$, we independently solve a local version of the source-free Maxwell's system subject to a set of twelve non-homogeneous Dirichlet linear boundary conditions (one for every edge of $\OmegakH$) given by
\begin{eqnarray}
\curl \E_l^{k} + \imath \omega \B_l^{k} &=& \vec{0}, \ \mbox{in} \  \OmegakH, \label{eq:em1_omegak} \\
\curl (\mu^{-1} \B_l^{k}) - \Sigma \E_l^{k} &=& \vec{0}, \ \mbox{in}  \ \OmegakH, \label{eq:me2_omegak} \\
\E_l^{k} \times \normalv &=& \bPhi_l \times \normalv, \  \mbox{on} \  \partial \OmegakH; \ \ l=1,\ldots,12, \nonumber \\ \label{eq:bc_omegak}
\end{eqnarray}    
where $\B_l^{k}, \E_l^{k}, \mu, \Sigma, \omega, \imath$ and $\normalv$ are defined as before in Section \ref{sec:mathModel}, $\partial \OmegakH$ denotes the boundary of $\OmegakH$, and each $\bPhi_l$ is a vector field that takes the value 1 along the tangential direction to the lth edge of $\OmegakH$ and decays linearly to 0 in the normal directions to the same edge (Figure \ref{fig:linearBC}). The set of twelve linear boundary conditions, $\{\bPhi_l\}_{l=1}^{12}$, form the natural basis functions for edge degrees of freedom (\cite{Monk2003}), hence they can be used to model general normal-linearly varying EM responses.   

% 2.1 Computation of basis functions
To solve the twelve local Maxwell's systems \eqref{eq:em1_omegak}-\eqref{eq:bc_omegak}, we use the fine mesh contained in $\OmegakH$ and the MFV method as described in Section \ref{sec:mfv}. That is, the set of discrete solutions can be obtained by solving twelve linear systems of the form \eqref{eq:system}. We refer to the set of discrete solutions $\{ \bfe_1^{k}, \ldots,\bfe_{12}^{k} \}$ as {\em multiscale basis functions}.

% 2.2 Properties of MS basis functions 
Note that using formulation \eqref{eq:em1_omegak}-\eqref{eq:bc_omegak} implies that $\E_l^k$ is oscillatory at the interior of the coarse cell $\OmegakH$, and it coincides with the natural basis functions $\{\bPhi_l\}_{l=1}^{12}$ at the boundary of $\OmegakH$, that is
\begin{equation} \label{eq:deltaij}
    \E_l^k \cdot \tangentv_{{\sf edge}_m} = \delta_{lm}; \ \ \ l,m = 1,\ldots,12,
\end{equation}
where $\tangentv_{{\sf edge}_m}$ is the unit tangent vector to the mth edge of $\partial \OmegakH$, and $\delta_{lm}$ is the Kronecker delta that takes the value 1 when $l=m$ and 0 otherwise.  
Naturally, the multiscale basis functions also satisfies these properties.  It follows that the tangential components of the multiscale basis functions are continuous at the boundaries of the coarse-mesh cells.  

% 2.3 Relationship with operator-induced interpolation
As shown in \cite{Hou1997,Efendiev2009,Haber2014b}, the multiscale basis functions can be arranged as the columns of a local coarse-to-fine interpolation matrix $\Pk$, i.e., $\Pk = \left[ \bfe_1^{k}, \ldots,\bfe_{12}^{k} \right]$, for the fine-mesh electric field in $\OmegakH$ (Step 2 in Figure \ref{fig:msfv}).  This type of interpolation is also known as operator-induced interpolation, which was originally developed for the diffusion equation with strongly discontinuous coefficients (cf. \cite{Alcouffe1981,Dendy1982}).  

% Step 3: Loop over all coarse cells to assemble global interpolation matrix using continuity
Once we have computed a local interpolation matrix for each coarse cell, the third step is to assemble a global coarse-to-fine interpolation matrix, $\PHh$ (Step 3 in Figure \ref{fig:msfv}). The continuity of the tangential components of the multiscale basis functions at the boundaries of the coarse cells is a necessary requirement for the proper assembly of $\PHh$ in this step (\cite{Efendiev2009}).

% 3.1 Multiscale methods can be done within a parallel framework
Observe that the calculations involved to compute the local interpolation matrices ($\Pk; \ k=1,\ldots,N$), are done locally inside each coarse cell independently of each oth\-er, hence they can perfectly be done in parallel. This great\-ly reduces the overhead time in constructing each $\Pk$ in practice.

% Step 4: construct coarse-mesh system
The fourth step is to use the global interpolation matrix $\PHh$ as a projection matrix within a Galerkin approach to construct a coarse-mesh version of the fine-mesh system \eqref{eq:mfvSystemNatBC} that is much cheaper to solve as follows
\begin{equation} \label{eq:msfvSystem}
  \PHh^{\top} {\bfA}^h({\pmb \Sigma}^h) \PHh {\bfe}^H =  \PHh^{\top} {\bf q}^h.
\end{equation}
The superscripts $H$ and $h$ denote dependency to the coarse and fine meshes, respectively, the vector ${\bfq}^h$ and the system matrix ${\bfA}^h({\pmb \Sigma}^h)$ are defined as in \eqref{eq:mfvSystemNatBC}, and $\bfe^H$ denotes the coarse-mesh electric field. 

% Recovering the fine-mesh solution from the coarse-mesh one
As shown in \cite{Haber2014b}, the fine-mesh electric field, $\bfe^h$, can be obtained from the solution to the coarse-mesh problem as follows 
\begin{equation}
{\bfe}^h = \PHh {\bfe}^H.
\end{equation}

In the above process we opt for the construction of the operator induced interpolation matrix $\PHh$.  However, it is possible to perform the computation cell-by-cell and to not generate the matrix.  While this approach has an advantage when considering storage, it requires the reconstruction of the fine-mesh field from the coarse-mesh one, which requires recomputing the basis function.  If the solution is needed only in a small number of cells, then this approach is preferred.

%
% Importance of choice of boundary conditions to construct basis functions, hence accurate sols 
The accuracy of the solutions obtained with multiscale FV/FE methods depends on the choice of boundary conditions used to construct the multiscale basis functions (second step outlined before) for each coarse cell. If these boundary conditions fail to reflect the effect of the underlying media heterogeneity contained by the coarse cell on the physical responses, multiscale procedures can have large errors (cf. \cite{Efendiev2009,Hou1997}).  

% Problem with linear boundary conditions and oversampling techniques for elliptic problems
Researchers in the field of multiscale methods for elliptic problems have noted that by choosing a set of linear boundary conditions for the construction of the multiscale basis functions, a mismatch between the exact solution and the discrete solution across the coarse cell boundary may be created, thus yielding to inaccurate solutions. The error analyses presented in \cite{Hou1997} and in \cite{Efendiev2009} demonstrate that the source of inaccuracy in the solution comes from resonance errors, that is errors that appear when the coarse-mesh size and the wavelength of the small scale oscillation of the media heterogeneity are similar.  A solution in such cases is to use oversampling techniques for the construction of the multiscale basis functions, (cf. \cite{Hou1997,Hou2003,Jenny2003,Efendiev2009,Hajibeygi2009}).  

% Transition
In the next section, we discuss the case where the choice of linear boundary conditions to construct the multiscale basis functions may yield inaccurate solutions using the MSFV method, and we develop an oversampling technique to fix this accuracy issue.

%------------------------------------------------------------
% SECTION
%------------------------------------------------------------
\subsection{The oversampling method} \label{sec:oversampling}

% Problem: linear boundary conditions in MSFV lead to inaccurate solutions
As discussed in the previous section, the MSFV method imposes linear boundary conditions to the local Max\-well's formulation \eqref{eq:em1_omegak}-\eqref{eq:bc_omegak} used to compute the multiscale basis functions inside each coarse-mesh cell. 
Note that by choosing linear boundary conditions for the multiscale basis functions, the MSFV method assumes that the tangential components of the electric field behaves linearly at the interfaces between coarse-mesh cells.
However, this assumption fails for the cases where the media contained by the coarse cells is highly heterogeneous, as it is well known that heterogeneous conductive media induces a non-linear and non-smooth behavior of the electric field (\cite{Ward1988}). In particular, when the heterogeneity is located close to the boundary of the coarse cell, the non-linear behavior of the electric field significantly violates the assumption of linear fields at the boundaries.
Hence, by imposing linear boundary conditions in such cases for the construction of the multiscale basis functions, the MSFV method creates a mismatch between the true and the multiscale solution across the coarse cell boundary. 
This mismatch yields to produce inaccurate solutions.  In this section, we propose an oversampling technique to overcome this difficulty.

% FIGURE
\begin{figure*}[bht!]
  \centering \includegraphics[width=1\textwidth]{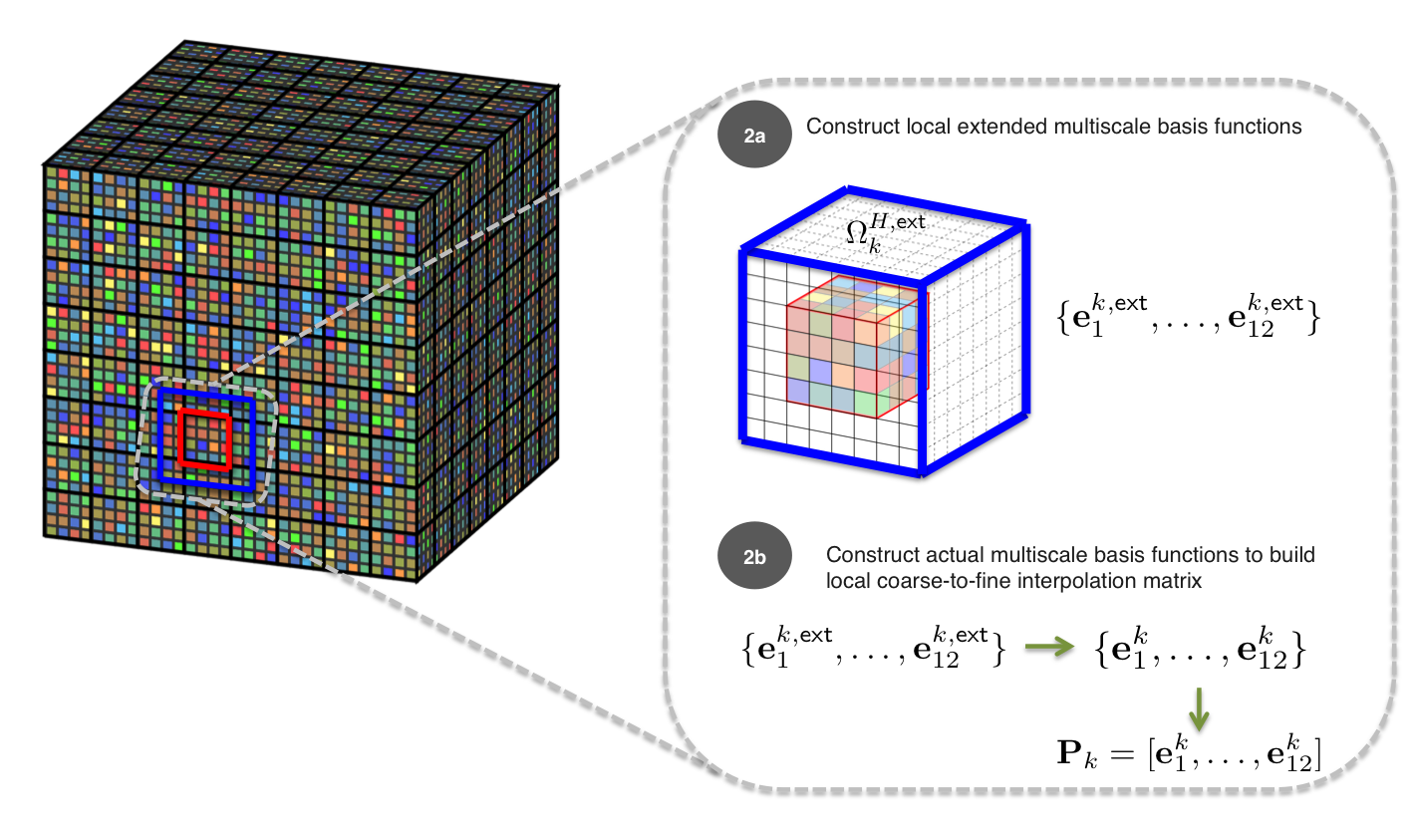}
  \caption{Schematic representation of the two main steps to implement the oversampling method. \label{fig:oversampling}}
\end{figure*}

% Alternative: oversampling methods - intuition
Oversampling methods are used to reduce boundary effects in the construction of the multiscale basis functions per single coarse-mesh cell (cf. \cite{Hou1997,Efendiev2009}). The main idea is to compute the multiscale basis functions using a local extended domain, and to use only the fine-mesh information at the interior of the cell to construct the multiscale basis functions.

% Transition to our oversampling technique
We now proceed to develop our oversampling technique.  To do so, we adapt the oversampling technique originally proposed by \cite{Hou1997} for elliptic problems with strongly discontinuous coefficients and for a nodal FE discretization, to apply to the MSFV method for EM modeling with edge variables (\cite{Haber2014a}) discussed in the previous section, which uses a staggered FV discretization.  

% Details oversampling
For a given coarse cell $\OmegakH$, the core idea behind our oversampling method consists of the following two steps (Figure \ref{fig:oversampling}): 

First, we compute multiscale basis functions using a {\em local extended domain}, $\OmegakHext$, which includes the coarse cell $\OmegakH$ and a neighborhood of fine cells around it. If the coarse cell $\OmegakH$ is at the boundary of the computational domain, then we only extend the local domain to where it is possible. To compute the multiscale basis functions in $\OmegakHext$, we formulate the twelve local Maxwell's systems as in \eqref{eq:em1_omegak}-\eqref{eq:bc_omegak}, but rather than using $\OmegakH$ as the local domain we use $\OmegakHext$, then we apply the MFV method as discussed in Section \ref{sec:mfv}. 
We refer to the set of discrete solutions $\{ \bfe_1^{k,{\sf ext}}, \ldots, \bfe_{12}^{k,{\sf ext}} \} $ as {\em extended multiscale basis functions}. 

% Define continuity of tangential components of multiscale basis functions
Second, we use the set of extended multiscale basis functions obtained in the previous step to compute the actual set of multiscale basis functions $\{ \bfe_1^k, \ldots, \bfe_{12}^k \}$ in $\OmegakH$.
Since the construction of the multiscale basis is done cell by cell, there is no guaranty that the tangential components of the multiscale basis functions are continuous at the boundary of the coarse cell $\OmegakH$. In order to mitigate this issue, we impose the following {\bf weak continuity condition} in the construction of to the  multiscale basis functions to warranty they will be weakly continuous along each shared boundary among immediate neighboring coarse cells,
\begin{eqnarray}
  \mathcal{A}_{{\sf edge}_m}\left(\E^k_l\right) := \frac{1}{L_{{\sf edge}_m}}\int_{{\sf edge}_m} \E^k_l \cdot \tangentv_{{\sf edge}_m} \  ds = \delta_{ml}; \nonumber \\  m,l = 1,\ldots,12, 
  \label{eq:deltaij_o} 
\end{eqnarray}
where $\E_l^k$ denotes the continuous form of the lth multiscale basis function $\bfe_l^k$, $L_{{\sf edge}_m}$ denotes the length of the mth edge of $\OmegakH$, $\tangentv_{{\sf edge}_m}$ denotes the unit tangent vector to the mth edge of $\OmegakH$, and $\delta_{ml}$ is the Kronecker delta.  That is, we take a `normalized average' of the multiscale basis functions at the boundary of the coarse cell.  This condition is equivalent to the definition of edge degrees of freedom of a staggered cell in the context of finite elements (cf. \cite{Jacobsson2007,Monk2003}). 
Note the difference with the continuity condition \eqref{eq:deltaij} imposed in the construction of the multiscale functions of the MSFV method without oversampling. 
% State condition numerically
Integrating numerically the continuity condition \eqref{eq:deltaij_o}, we can express it as
\begin{equation} \label{eq:bc_msbf}
\mA_{{\sf edge}_m}\left(\bfe^k_l\right) \approx {\bf v}^{\top}_{{\sf edge}_m}{\bfe^k_l} = \delta_{ml}; \ \ m,l = 1,\ldots,12, 
\end{equation}
where ${\bf v}_{{\sf edge}_m}$ is the vector that computes the normalized line integral along the mth edge of $\OmegakH$. 

% Intro for procedure of oversampling
Using \eqref{eq:bc_msbf} and following the main lines given in the oversampling technique proposed by \cite{Hou1997}, we continue the development of our oversampling technique by showing how to compute $\{ \bfe_1^k, \ldots, \bfe_{12}^k \}$ from $\{ \bfe_1^{k,{\sf ext}}, \ldots, \bfe_{12}^{k,{\sf ext}} \}$ in detail.

% Step 1: 
We begin by expressing the jth multiscale basis function, $\bfe_j^k$, as a linear combination of the set of extended basis functions as follows
\begin{equation}
  \bfe_j^k = \sum_{l=1}^{12} c_{l,j} \, \bfe_l^{\sf ext} = [\bfe_1^{k,{\sf ext}},\ldots,\bfe_{12}^{k,{\sf ext}}] \, {\bf c}_j; \ \ j = 1,\ldots,12, \label{eq:lcOversampling}
\end{equation}
where ${\bf c}_j = [c_{1,j},\ldots,c_{12,j}]^{\top}$ are coefficients to be determined.  Now, to determine uniquely  such coefficients, we apply condition \eqref{eq:bc_msbf} to \eqref{eq:lcOversampling} which yields to the system of equations 
\begin{equation}
    \left[  \begin{array}{c c c c} 
        \mA_{\sf edge_1}(\bfe_1^{k,{\sf ext}})  & \mA_{\sf edge_1}(\bfe_2^{k,{\sf ext}})  & \dotsc & \mA_{\sf edge_{1}}(\bfe_{12}^{k,{\sf ext}})  \\
        \mA_{\sf edge_2}(\bfe_1^{k,{\sf ext}})  & \mA_{\sf edge_2}(\bfe_2^{k,{\sf ext}})  & \dotsc & \mA_{\sf edge_{2}}(\bfe_{12}^{k,{\sf ext}})  \\
        \vdots   & \vdots   & \ddots & \vdots    \\
        \mA_{\sf edge_{12}}(\bfe_{1}^{k,{\sf ext}}) & \mA_{\sf edge_{12}}(\bfe_{2}^{k,{\sf ext}}) & \dotsc & \mA_{\sf edge_{12}}(\bfe_{12}^{k,{\sf ext}}) \\
          \end{array} 
    \right] 
\bfC =  \bfI_{12 \times 12}, \label{eq:lcCondO}
\end{equation}
where $\bfC = [{\bf c}_1,\ldots,{\bf c}_{12}]$ and ${\bf I}_{12x12}$ denotes the 12 by 12 identity matrix.  Combining equations \eqref{eq:bc_msbf}, \eqref{eq:lcOversampling} and \eqref{eq:lcCondO}, we obtain the expression for the desired coefficients, that is
\begin{equation}
    \bfC=   
    {\left[   \begin{array}{c c c c} 
        {\bf v}^{\top}_{{\sf edge}_1}\bfe_1^{k,{\sf ext}}  & {\bf v}^{\top}_{{\sf edge}_1}\bfe_2^{k,{\sf ext}}  & \dotsc & {\bf v}^{\top}_{{\sf edge}_{1}}\bfe_{12}^{k,{\sf ext}}  \\
        {\bf v}^{\top}_{{\sf edge}_2}\bfe_1^{k,{\sf ext}}  & {\bf v}^{\top}_{{\sf edge}_2}\bfe_2^{k,{\sf ext}}  & \dotsc & {\bf v}^{\top}_{{\sf edge}_{2}}\bfe_{12}^{k,{\sf ext}}  \\
        \vdots   & \vdots   & \ddots & \vdots    \\
        {\bf v}^{\top}_{{\sf edge}_{12}}\bfe_1^{k,{\sf ext}}  & {\bf v}^{\top}_{{\sf edge}_{12}}\bfe_2^{k,{\sf ext}}  & \dotsc & {\bf v}^{\top}_{{\sf edge}_{12}}\bfe_{12}^{k,{\sf ext}} \\
          \end{array} 
    \right]}^{-1}. \label{coeffsO}    
\end{equation}
%

% Connect with general workflow of MSFV
After we construct the multiscale basis functions \\ $\{ \bfe_1^k, \ldots, \bfe_{12}^k \}$ using our oversampling technique, we continue to follow the procedure for the MSFV method (Figure \ref{fig:msfv}) to compute the solution.  That is, the multiscale basis functions $\{ \bfe_1^k, \ldots, \bfe_{12}^k \}$ enable the use of the local interpolation matrix $\Pk$, given by $\Pk = \left[\bfe_1^k, \ldots, \bfe_{12}^k \right]$, within the assembly of the global coarse-to-fine interpolation matrix $\PHh$.  The interpolation matrix $\PHh$ is then used within a Galerkin formulation to obtain the coarse-mesh system \eqref{eq:msfvSystem}, which we ultimately solve.

%------------------------------------------------------------
% SIMULATIONS
%------------------------------------------------------------
\section{Numerical results} \label{sec:numres}

% Section overview:
In this section, we demonstrate the accuracy and computational performance of our proposed oversampling technique for the multiscale finite volume (MSFV+O) meth\-od using a synthetic 3D model of a mineral deposit in a complex geological setting. 

%------------------------------------------------------------
% \subsection{Example 2} \label{sec:exa2}

% Left: reference model, right: coarse model
\begin{figure}[ht]
  \centering
  \begin{tabular}[c]{c c}
    \begin{subfigure}[c]{0.46\textwidth}
      \includegraphics[width=\textwidth]{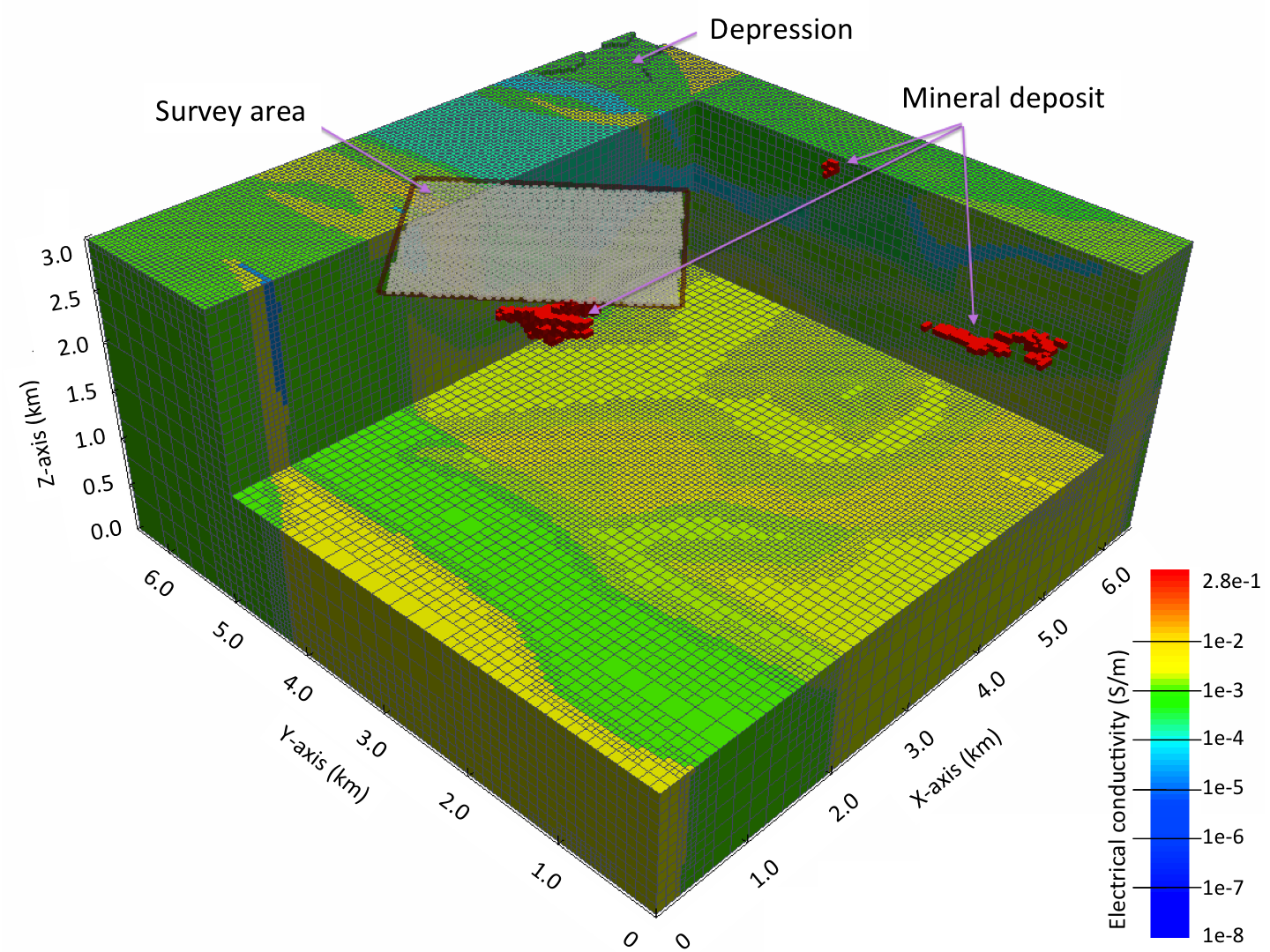}
      \caption{}
      \label{fig:refModel}
    \end{subfigure} &
    \begin{subfigure}[c]{0.46\textwidth}
      \includegraphics[width=\textwidth]{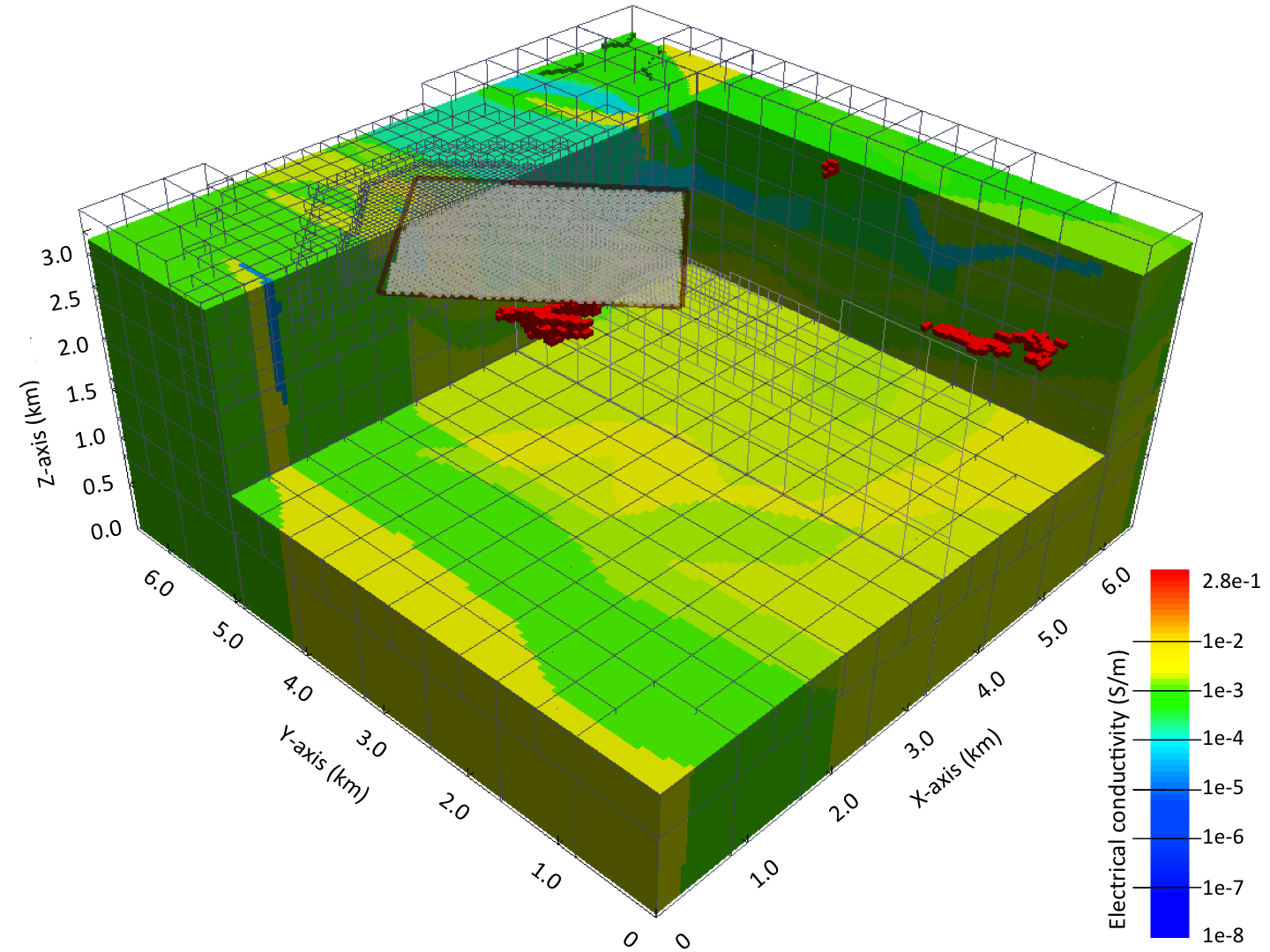}
      \caption{}
      \label{fig:coarsedModel}
    \end{subfigure}
  \end{tabular}    
    \caption{Subsurface part of the synthetic electrical conductivity model and large-loop EM survey setup. (a) Model discretized on a fine OcTree mesh (546,295 cells). The whole conductivity model varies over eight orders of magnitude. (b) Model with an overlaying coarse OcTree mesh (60,656 cells). The coarse OcTree mesh maintains the same cell size as the fine mesh in the survey area and gradually increases the cell size for the rest of the domain.}
    % TD: % Change figures for slices?  
    \label{fig:condModel}
\end{figure}

% Describe generation and composition of synthetic conductivity model and survey setup
For this example, we construct a synthetic electrical conductivity model based on the inversion results of field measurements over the Canadian Lalor mine obtained by \cite{Yang2014}.  The Lalor mine targets a large zinc-gold-copper deposit that has been the subject of several EM surveys.  
The synthetic conductivity model, shown in Figure \ref{fig:refModel}, has an area with non-flat topography and extends from 0 to 6.5 km along the $x$, $y$ and $z$ directions, respectively. The model comprises air and the subsurface that is composed of 35 geologic units.  The unit with the largest conductivity value represents the mineral deposit, which is composed of three bodies. We assume a conductivity of $10^{-8}$ S/m in the air. The subsurface conductivity values range from $1.96 \times 10^{-5}$ S/m to $0.28$ S/m. 
We consider a large-loop EM survey for this example, where we use a rectangular transmitter loop with dimensions 2 km $\times$ 3 km, operating at the frequencies of 1, 10, 20, 40, 100, 200 and 400 Hz. The transmitter is placed on the Earth's surface and it is centered above the largest body of the mineral deposit, as shown in Figure \ref{fig:refModel}.  Inside the loop, we place a uniform grid of receivers that measure the three components of the magnetic flux ($\B = [B^x, B^y, B^z]^\top$).  The receivers are separated by 50 m along the $x$ and $y$ directions, respectively. 
To reduce the effect of the imposed natural boundary conditions \eqref{eq:naturalBC}, we embed the survey area into a much larger computational domain, which replaces the true decay of the fields towards infinity (Figure~\ref{fig:refModel}). 

% Purpose of this simulation
Our aim is to estimate the {\em secondary magnetic flux} induced by the mineral deposit in the survey area.
For this purpose, we simulate two sets of the magnetic flux data for each frequency. The first data set considers the conductivity model including all geologic units, and the second data set excludes the mineral deposit from the original conductivity model.
The secondary magnetic flux induced by the mineral deposit, denoted as $\Delta \B$, is then computed by subtracting the two data sets. 

\begin{figure}[htb!]
  \centering \includegraphics[width=0.8\textwidth]{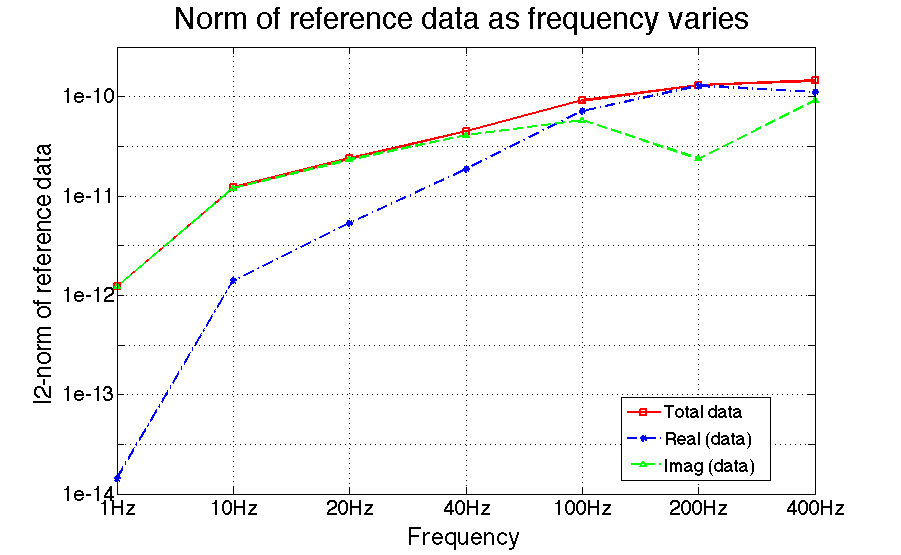}
  \caption{$l_2$ norm of total, real and imaginary parts of the reference data $\Delta \B$ per frequency.} 
  \label{fig:normRefSol}
\end{figure}

\begin{figure*}[ht!]
  \centering
  \begin{tabular}[c]{cc}
    \begin{subfigure}[c]{0.40\textwidth}
      \includegraphics[width=\textwidth]{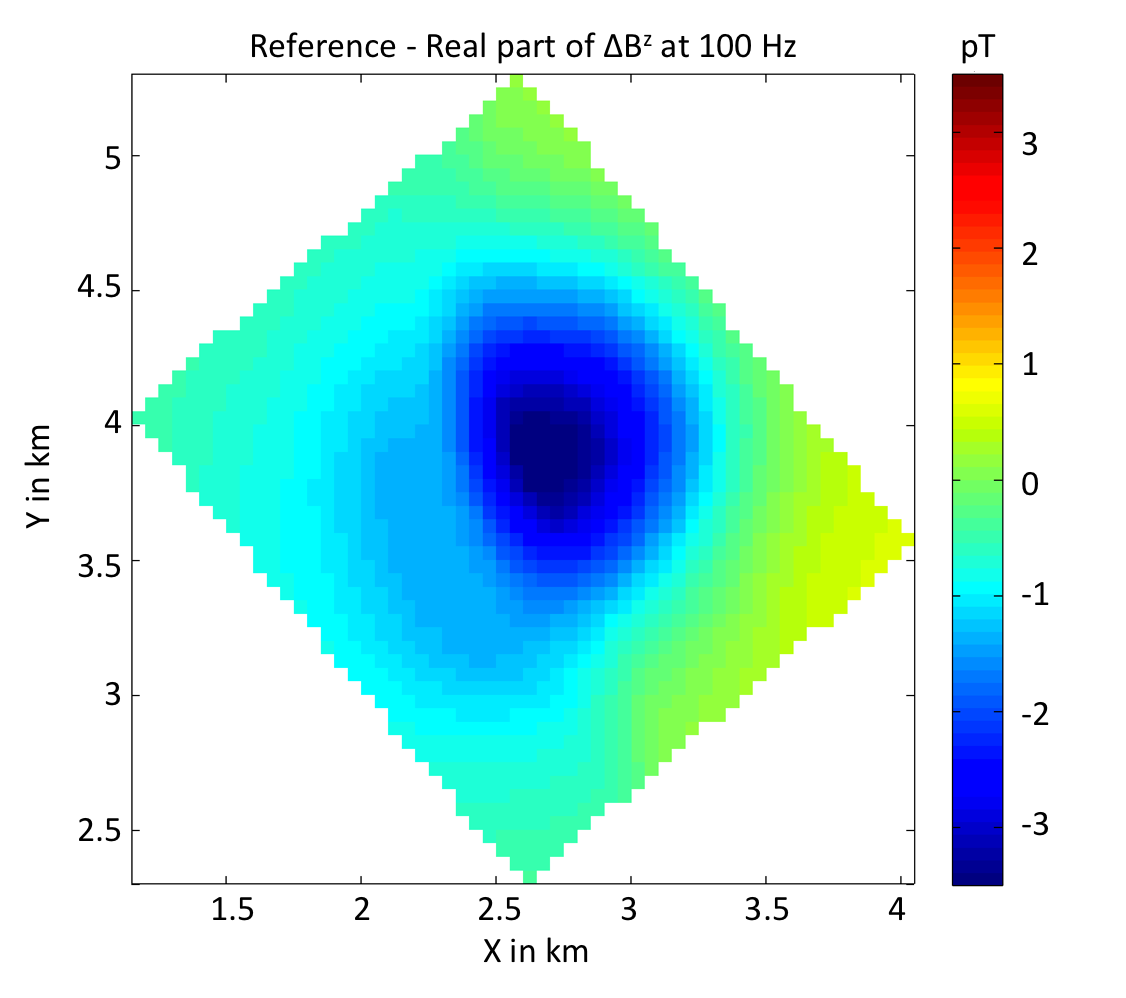}
      \caption{}
      \label{fig:rRef}
    \end{subfigure}&
    \begin{subfigure}[c]{0.40\textwidth}
      \includegraphics[width=\textwidth]{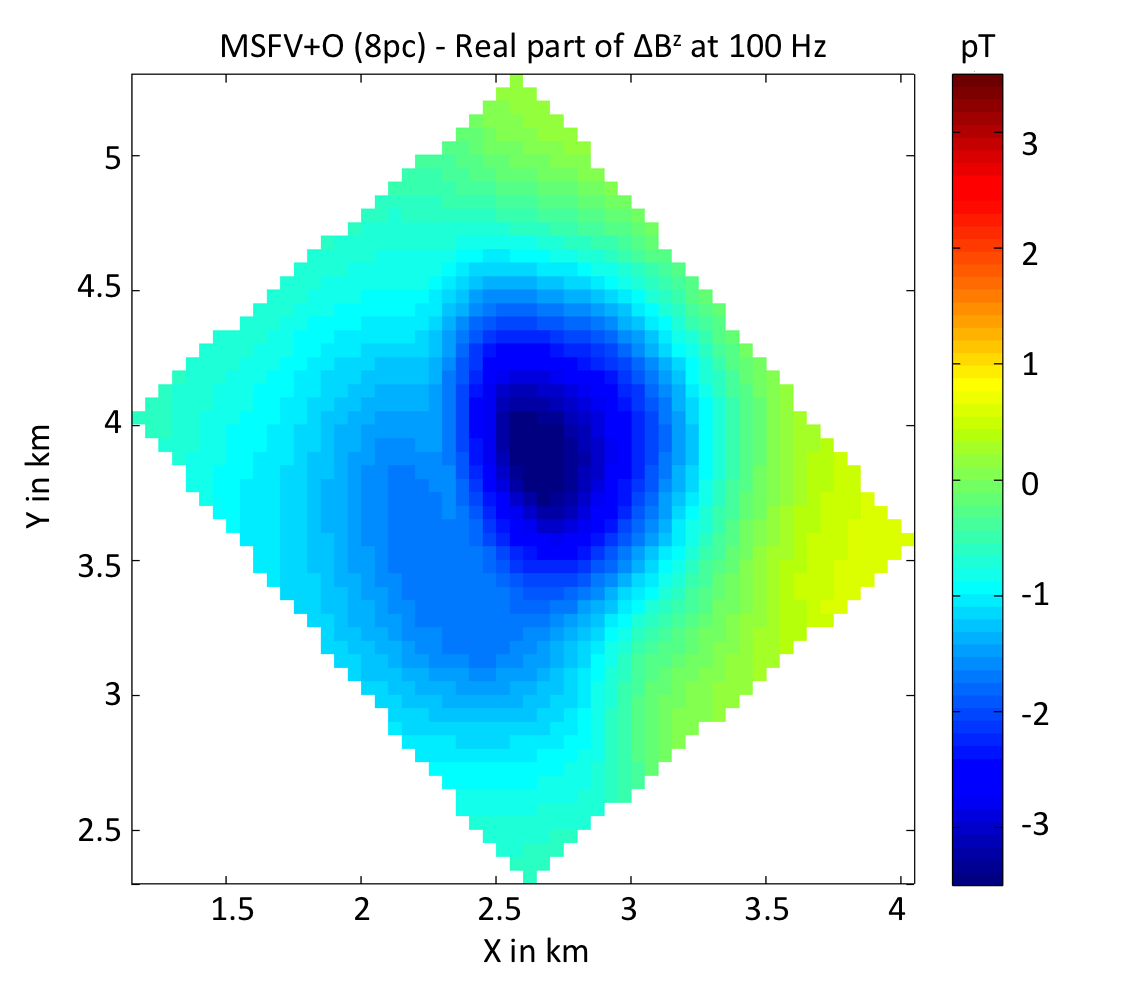}
      \caption{}
      \label{fig:rMSFVO}
    \end{subfigure} \\
    \begin{subfigure}[c]{0.40\textwidth}
      \includegraphics[width=\textwidth]{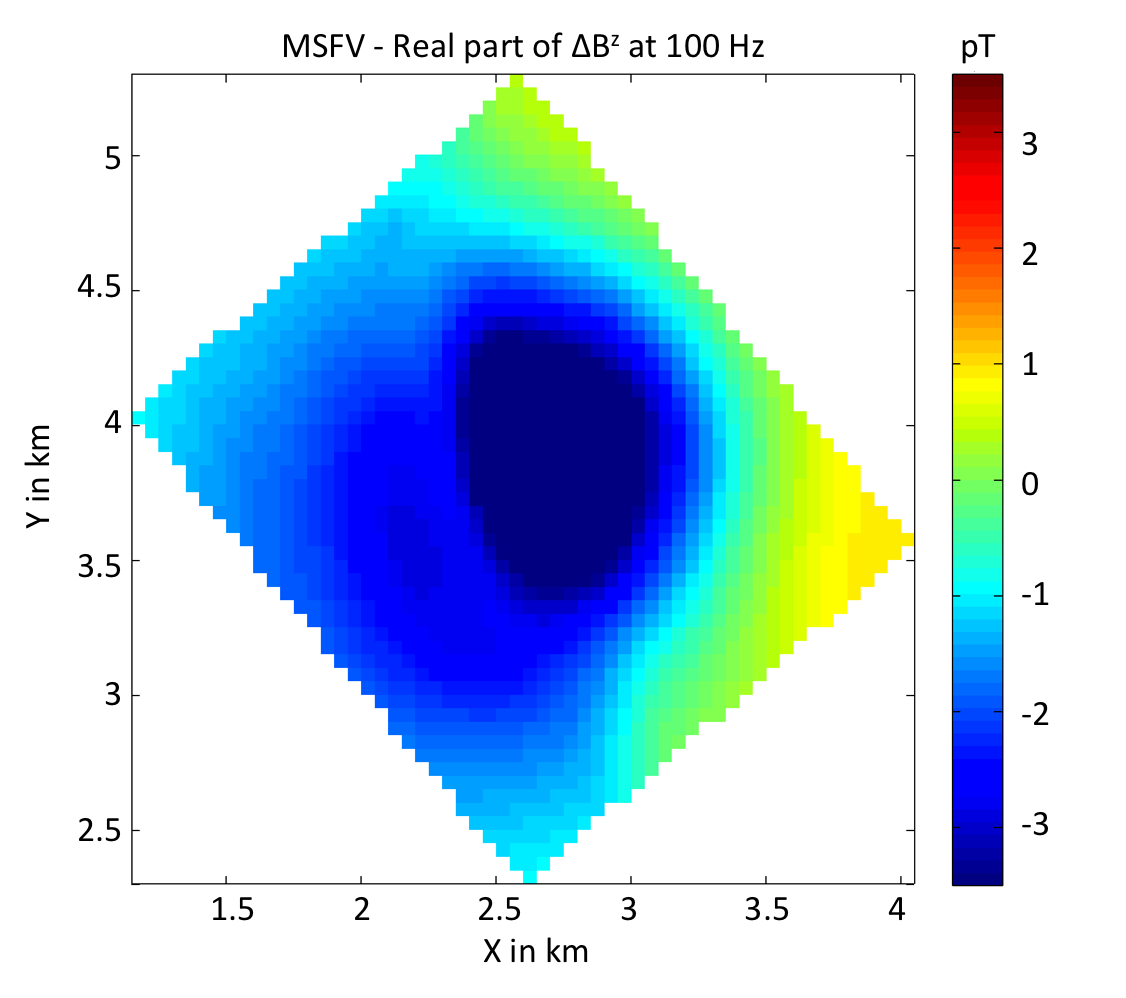}
      \caption{}
      \label{fig:rMSFV}
    \end{subfigure} & 
    \begin{subfigure}[c]{0.40\textwidth}
      \includegraphics[width=\textwidth]{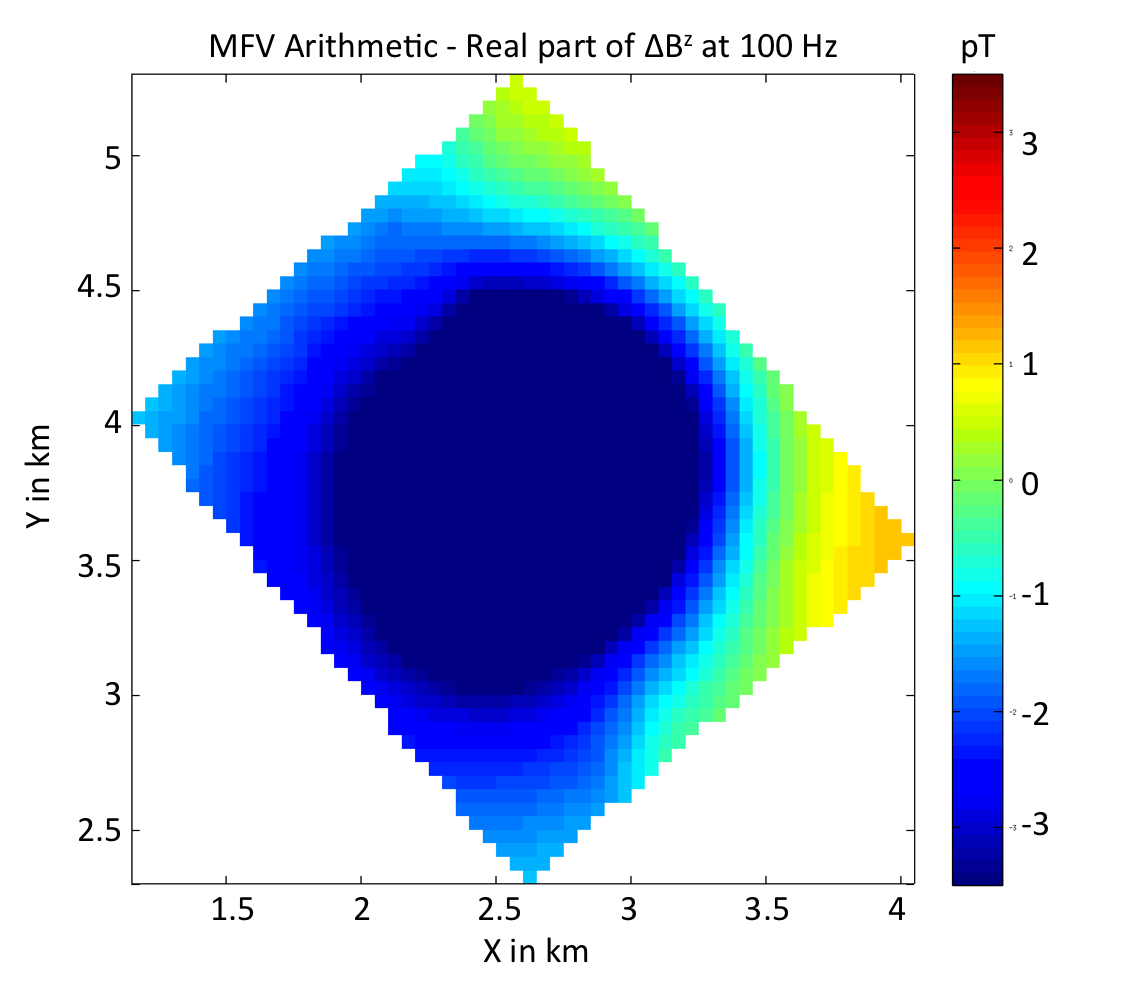}
      \caption{}
      \label{fig:rArith}
  \end{subfigure} \\
    \begin{subfigure}[c]{0.40\textwidth}
      \includegraphics[width=\textwidth]{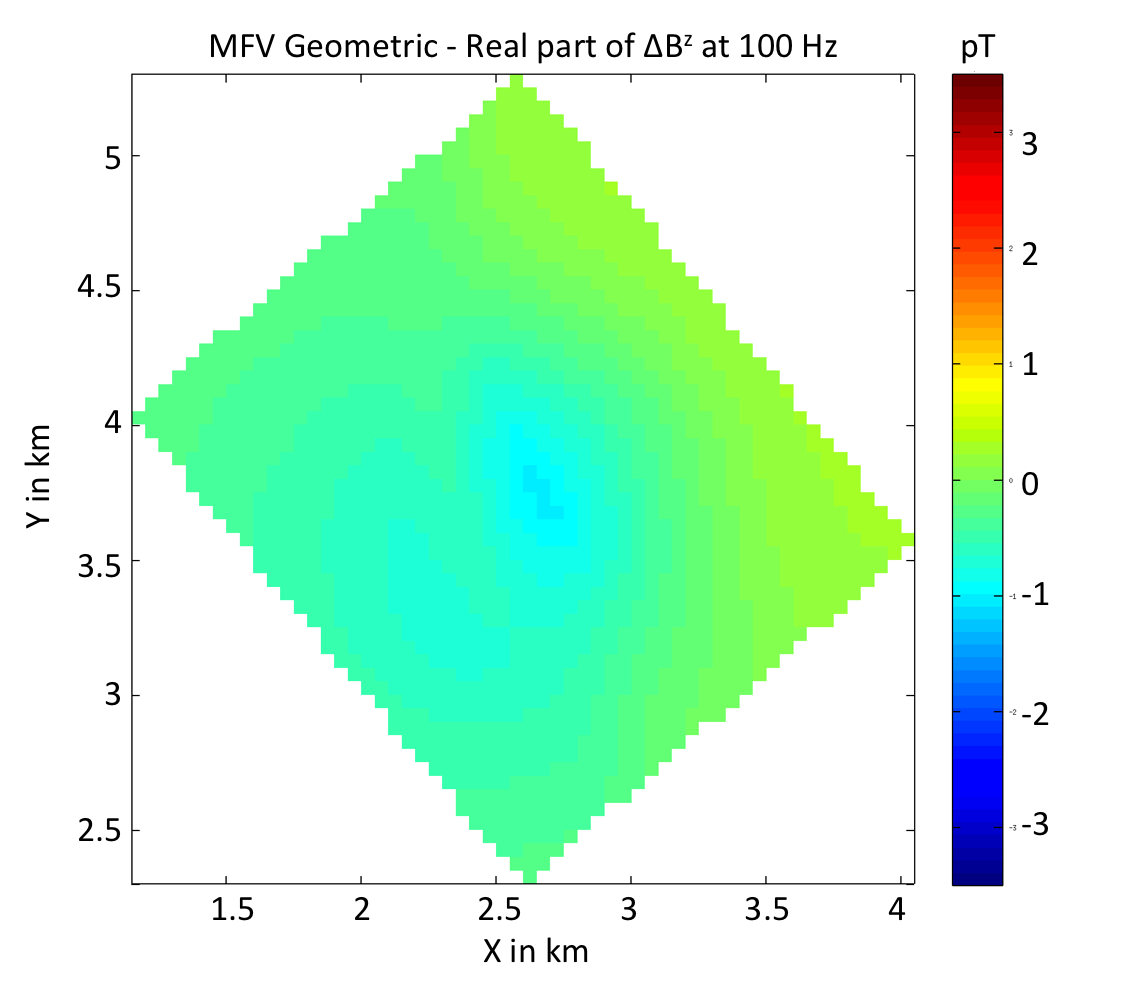}
      \caption{}
      \label{fig:rGeom}
    \end{subfigure}&
    \begin{subfigure}[c]{0.40\textwidth}
      \includegraphics[width=\textwidth]{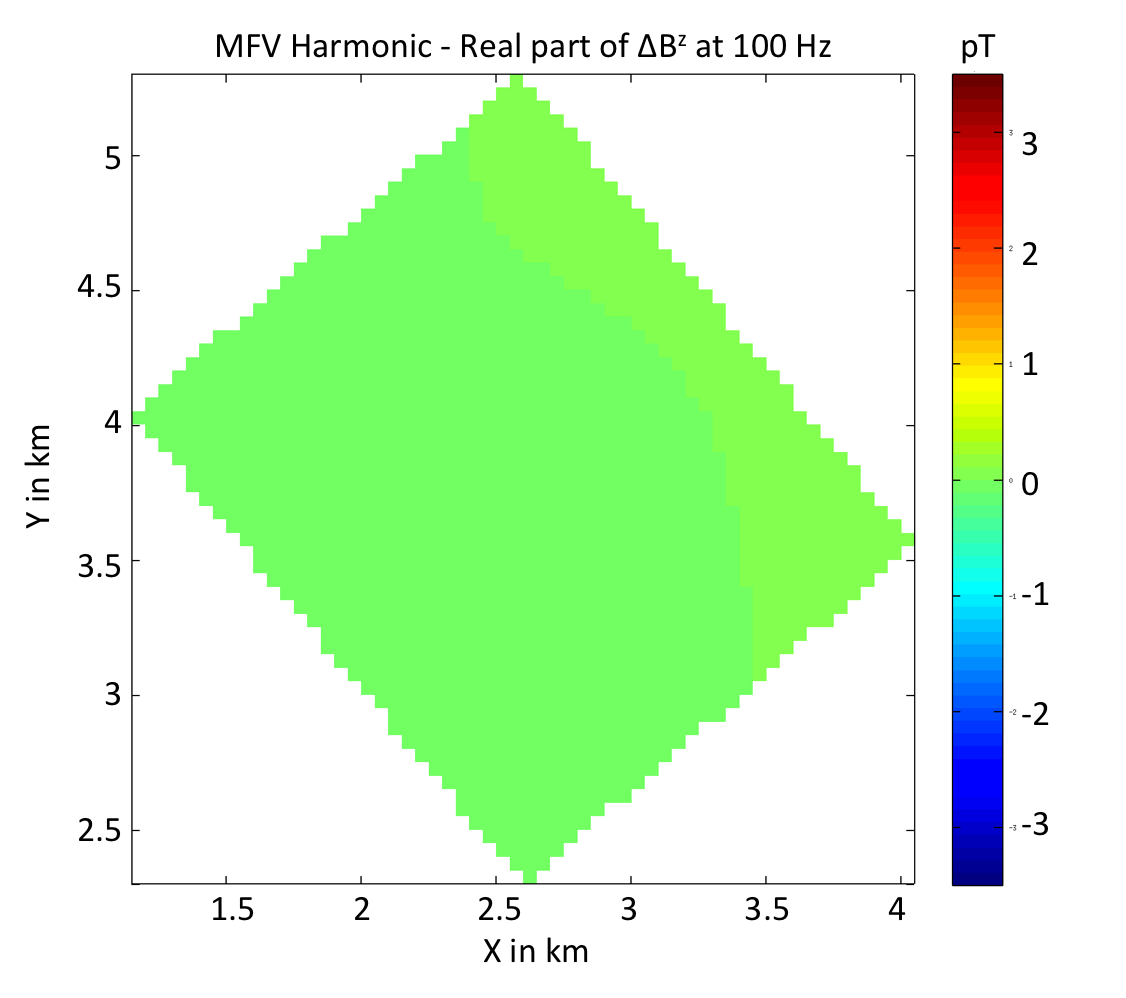}
      \caption{}
      \label{fig:rHarm}
  \end{subfigure} \\
  \end{tabular}    
    \caption{Real parts of the z-component of the secondary magnetic fluxes induced by the mineral deposit, $\Delta B^z$, for our large-loop EM survey at 100 Hz. (a) reference solution computed using the MFV method on the fine OcTree mesh with 546,295 cells. (b) and (c): results using the MSFV+O (with 8 padding cells) and MSFV methods on the coarse OcTree mesh with 60,656 cells, respectively.  (d), (e) and (f): results using MFV with the conductivity model homogenized using arithmetic, geometric and harmonic averaging on the coarse OcTree mesh, respectively.  All results are shown in picoteslas (pT) and plotted using the same color scale. \label{fig:realResults}}
\end{figure*}

\begin{figure*}[ht!]
  \centering
  \begin{tabular}[c]{cc}
    \begin{subfigure}[c]{0.40\textwidth}
      \includegraphics[width=\textwidth]{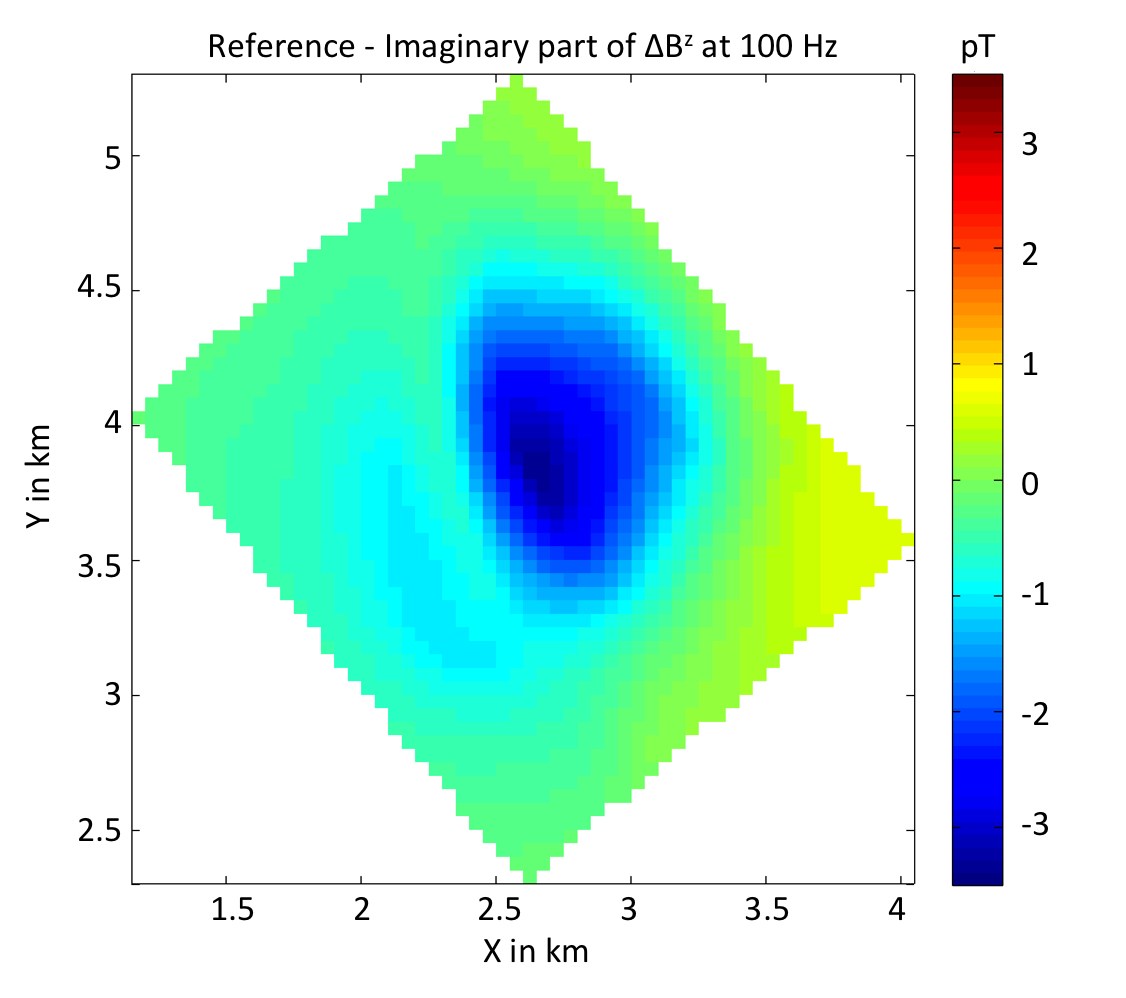}
      \caption{}
      \label{fig:iRef}
    \end{subfigure}&
    \begin{subfigure}[c]{0.40\textwidth}
      \includegraphics[width=\textwidth]{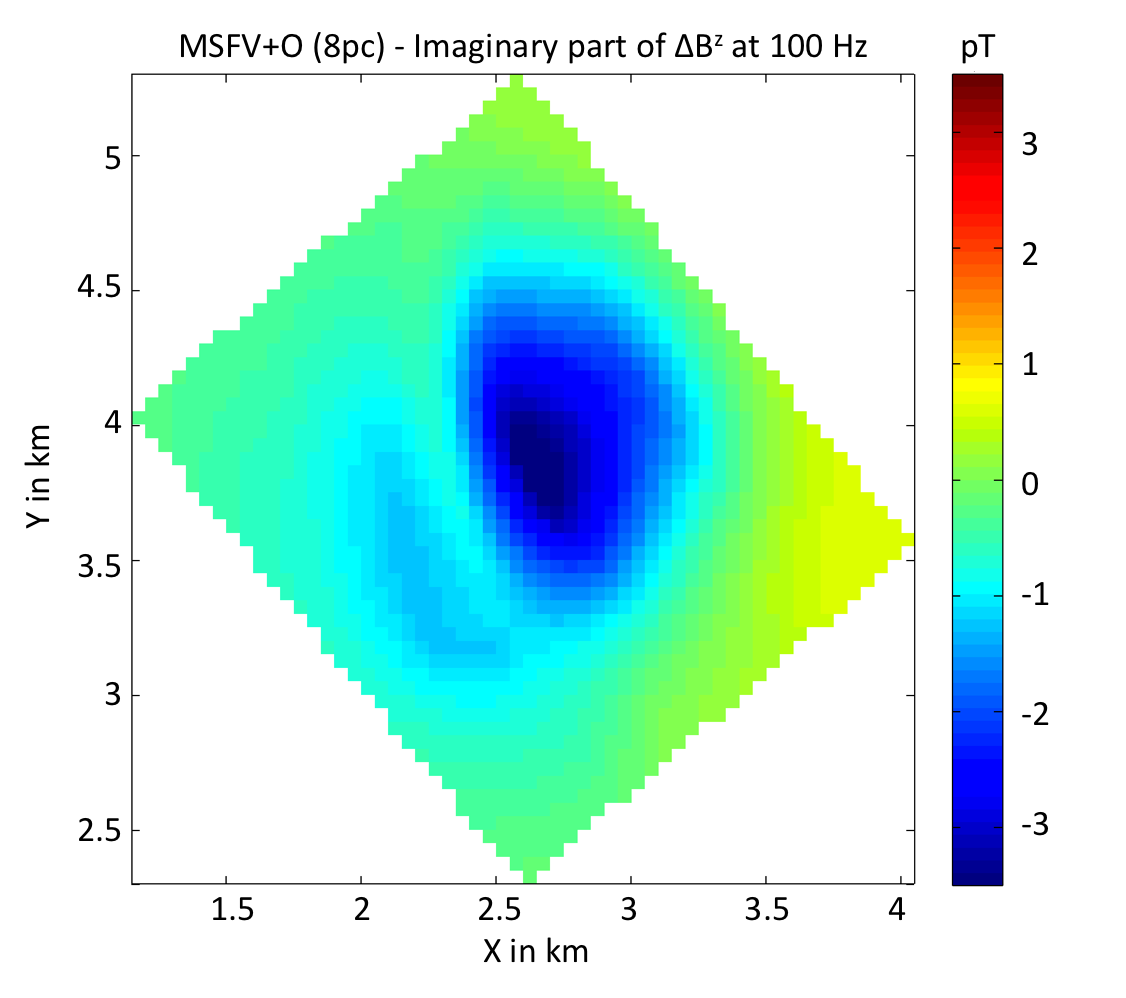}
      \caption{}
      \label{fig:iMSFVO}
    \end{subfigure} \\
    \begin{subfigure}[c]{0.40\textwidth}
      \includegraphics[width=\textwidth]{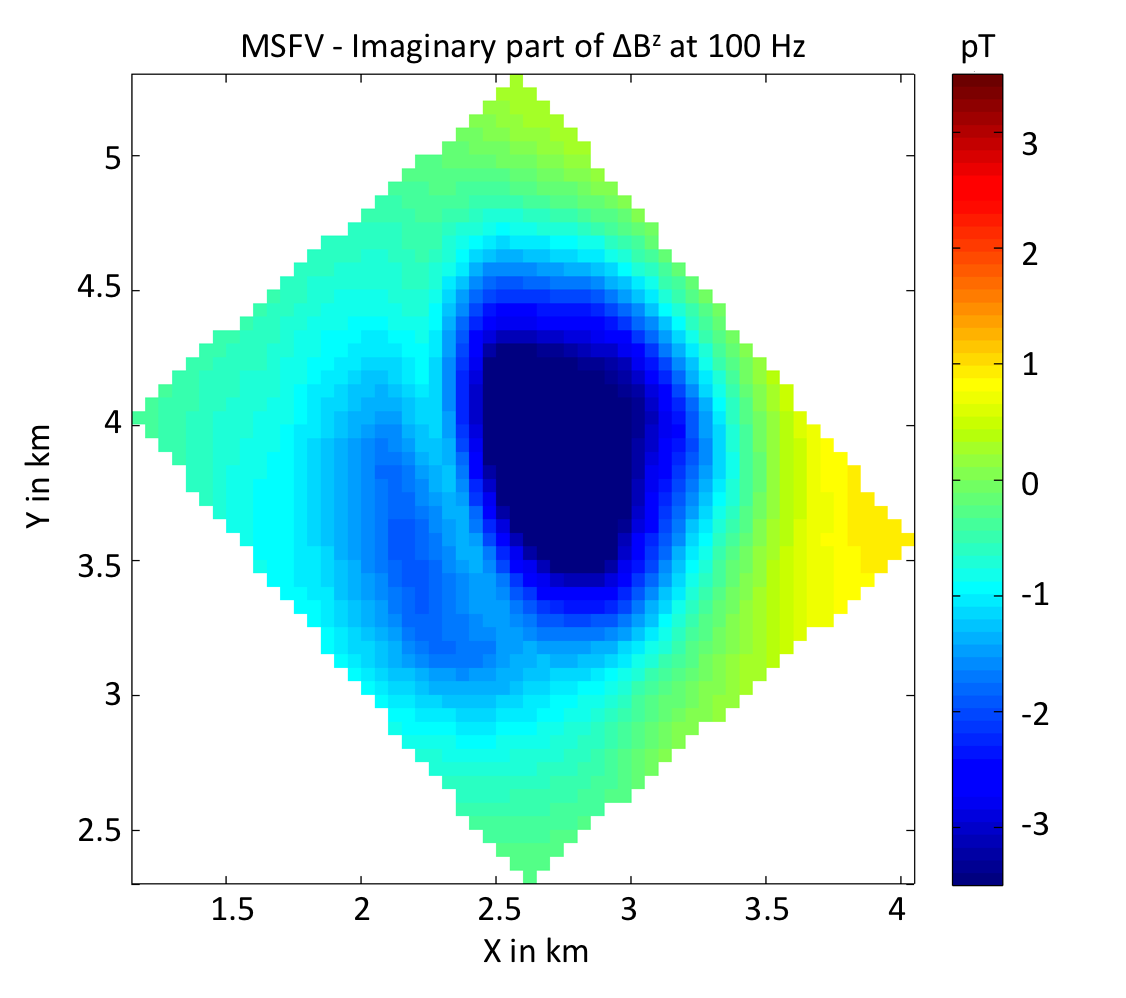}
      \caption{}
      \label{fig:iMSFV}
    \end{subfigure} &
    \begin{subfigure}[c]{0.40\textwidth}
      \includegraphics[width=\textwidth]{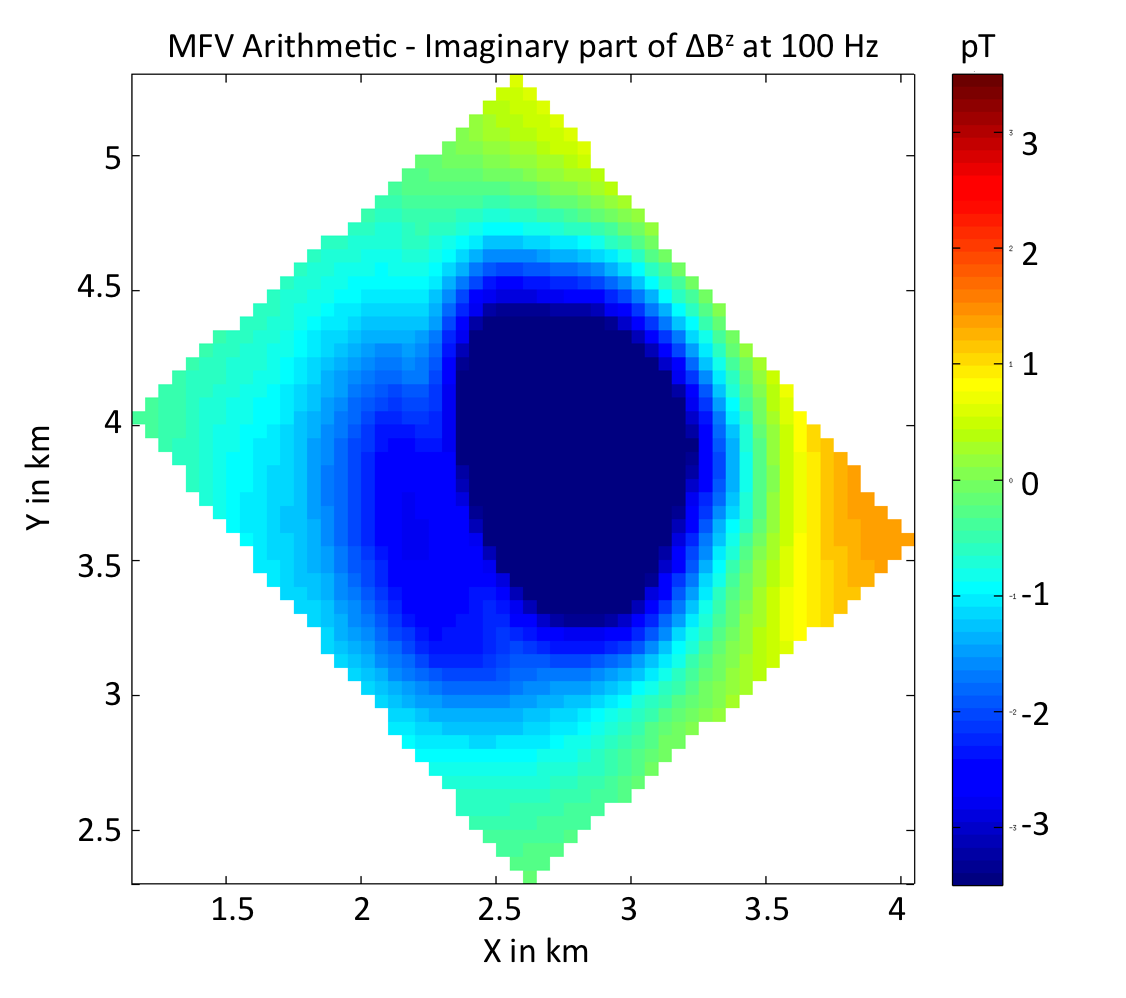}
      \caption{}
      \label{fig:iArith}
  \end{subfigure} \\
    \begin{subfigure}[c]{0.40\textwidth}
      \includegraphics[width=\textwidth]{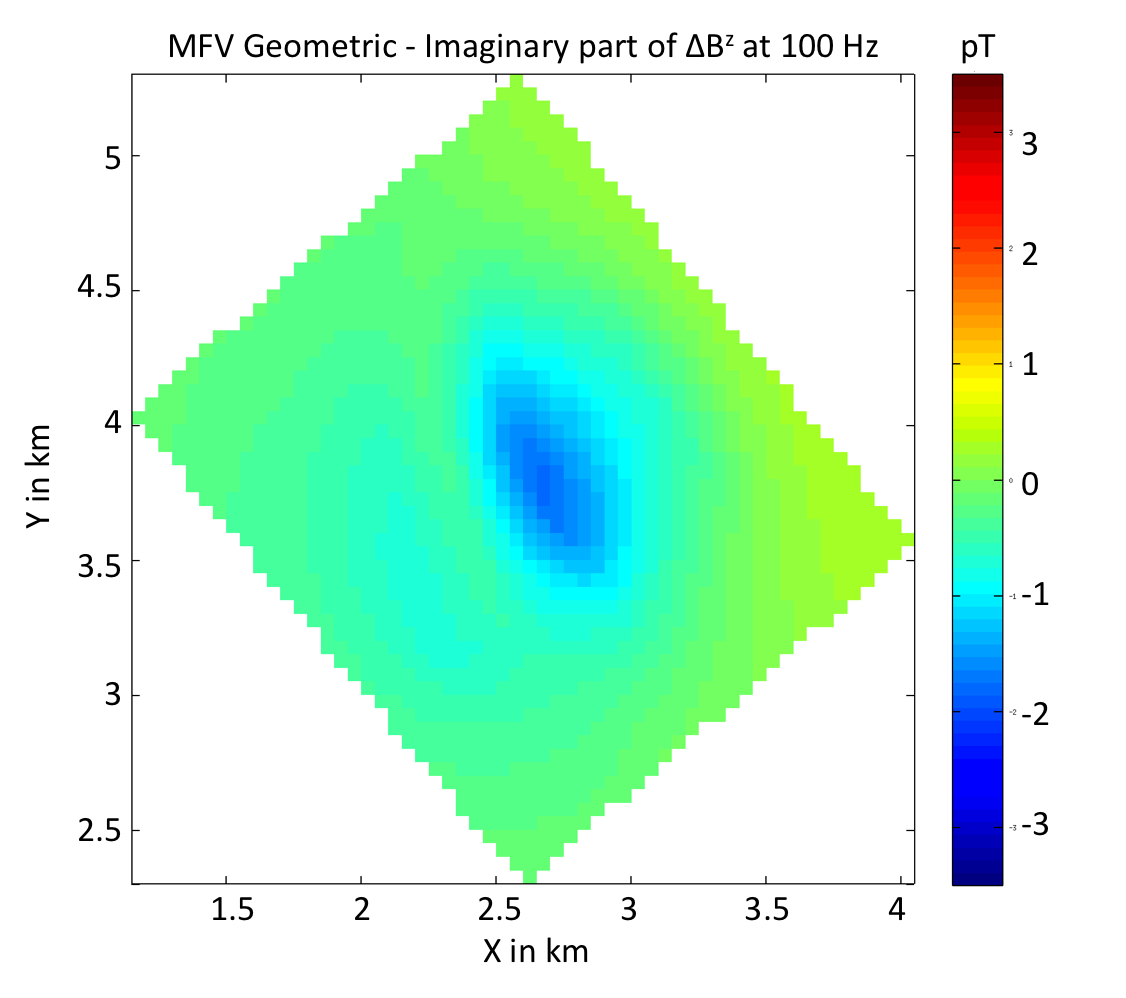}
      \caption{}
      \label{fig:iGeom}
    \end{subfigure}&
    \begin{subfigure}[c]{0.40\textwidth}
      \includegraphics[width=\textwidth]{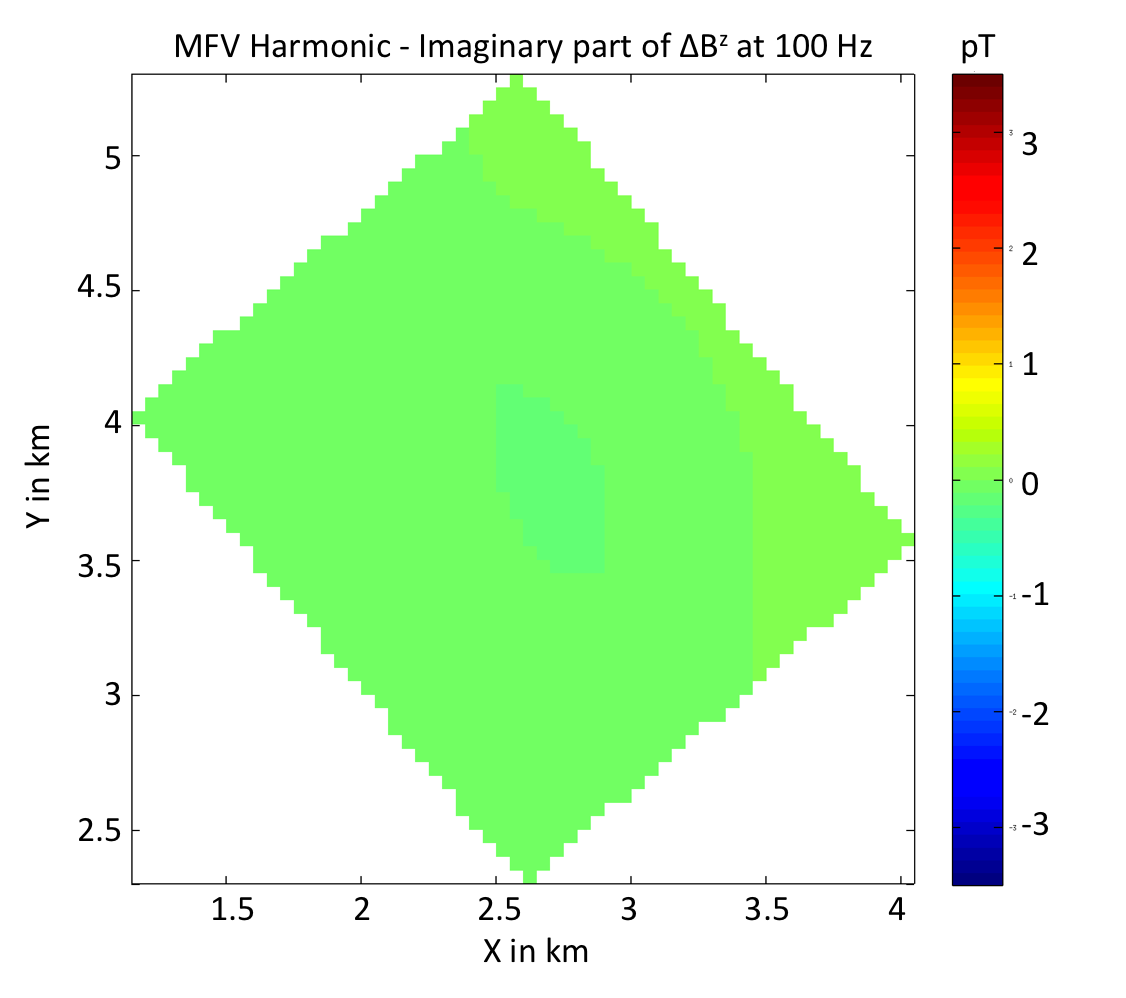}
      \caption{}
      \label{fig:iHarm}
  \end{subfigure} \\
  \end{tabular}    
    \caption{Imaginary parts of the z-component of the secondary magnetic fluxes induced by the mineral deposit, $\Delta B^z$, for our large-loop EM survey at 100 Hz. (a) reference solution computed using the MFV method on the fine OcTree mesh with 546,295 cells. (b) and (c): results using the MSFV+O (with 8 padding cells) and MSFV methods on the coarse OcTree mesh with 60,656 cells, respectively.  (d), (e) and (f): results using MFV with the conductivity model homogenized using arithmetic, geometric and harmonic averaging on the coarse OcTree mesh, respectively.  All results are shown in picoteslas (pT) and plotted using the same color scale. \label{fig:imagResults}}
\end{figure*}

% Reference solution: generation details of fine OcTree mesh, MFV, MUMPS (time to solve system and number of DOF)
To compute a reference solution, first we discretize our electrical conductivity model using a staggered fine OcTree mesh, and then we apply the MFV method as discussed in \cite{Haber2007a,Horesh2011}. 
We base the estimate for the proper cell sizes of the mesh on the skin depth (\cite{Ward1988}).  Practical experience on mesh design for EM problems reported in \cite{Haber2014} suggests that the smallest cell size in the mesh should be a quarter the minimum skin depth. 
We consider the largest background conductivity value (4.5 $\times 10^{-3}$ S/m) and calculate skin depths of 7,461, 2,359, 1,668, 1,180, 746, 527 and 373 m for the 1, 10, 20, 40, 100, 200 and 400 Hz frequencies, respectively.  
Hence, we use cells of size $(50 \ \mbox{m})^{3}$ within the survey area and at the interfaces of the model where the conductivity varies, the rest of the domain is padded with gradually increasing OcTree cells. This is illustrated in Figure \ref{fig:refModel}. This mesh has 546,295 cells.   
Using the MFV method on the fine OcTree mesh yields systems with roughly 1.5 millions degrees of freedom (DOF) which we solve using the parallel sparse direct solver MUMPS (\cite{Amestoy2001}). 
The average computation time per single simulation is 712 s on a two hexa-core Intel Xeon X5660 CPUs at 2.8 Hz, 64 GB shared RAM using MATLAB.
Figure \ref{fig:normRefSol} shows the $l_2$ norm of the total, real and imaginary parts of $\Delta \B$ for each frequency considered.  The real and imaginary parts of the results obtained for the z-component of $\Delta \B$, denoted as $\Delta B^z$, at 100 Hz are shown in Figure~\ref{fig:rRef} and Figure~\ref{fig:iRef}, respectively. 

% Setup MSFV+O method: details generation of coarse mesh, size of local extended domain
In order to use the MSFV+O method introduced in Section \ref{sec:msfvo}, we need to choose a suitable coarse mesh to discretize the conductivity model and the size of the local extended domain to compute its corresponding projection matrix. 
% Generation of the coarse OcTree mesh and reason to choose this mesh -  how do we know this coarse mesh is appropriate?
We consider a coarser OcTree mesh nested in the fine OcTree mesh previously described. The coarser OcTree mesh is designed to maintain the fine-mesh resolution of $(50 \ \mbox{m})^3$ inside the survey area, whereas the rest of the domain is filled with gradually increasing coarser cells. 
Figure \ref{fig:coarsedModel} shows the large-loop survey setup using this coarse mesh. In total, this mesh contains 60,656 cells. 
To analyze the performance of our MSFV+O meth\-od for coarse OcTree meshes, we do not refine the mesh outside the survey area where a large conductivity contrast is present in the model. For example, this mesh discretizes the mineral deposit with cells of size $(200 \ \mbox{m})^3$ and $(400 \ \mbox{m})^3$, and the non-flat topography (depression) with cells of size $(400 \ \mbox{m})^3$ and $(800 \ \mbox{m})^3$ (Figure \ref{fig:coarsedModel}). 
% Point out the method will be challenged under excessive coarsening
We note that the simulations for the frequencies of 200 and 400 Hz may be particularly challenging for our MSFV+O method as the coarsening in those areas can be considered extreme due to the cell size is in the order of the skin depth.
% Size of extended local domain
Next, to investigate the effect of the size of the local extended domain, i.e., the number of fine-mesh padding cells by which we extend every coarse cell, on the resulting magnetic flux data, we pad the coarse cell using 2, 4, and 8 fine cells. 
Each fine padding cell is of size $(50 \ \mbox{m})^3$.
The chosen local extended domain sizes correspond to extending each $(200 \ \mbox{m})^3$-coarse cell by half, one and two coarse cells.  The $(200 \ \mbox{m})^3$-coarse cells are the majority of the coarse cells where the largest conductivity contrast happen in this setting (Figure \ref{fig:coarsedModel}).

% Application of MSFV method with and without oversampling, size of reduced system, use of MUMPS, times for each case
Applying MSFV and MSFV+O on the coarse mesh shown in Figure~\ref{fig:coarsedModel}, we obtain reduced linear systems with 169,892 DOF, which are also solved using MUMPS. 
When using MSFV+O the total average run times per single simulation for extended domain sizes of 2, 4 and 8 padding cells are 160, 446 and 3,646 s, respectively, on the same machine. The real and imaginary parts of $\Delta B^z$ at 100 Hz for an extended domain size of 8 padding cells are shown in Figure~\ref{fig:rMSFVO} and Figure~\ref{fig:iMSFVO}, respectively.
% Comparison with MSFV method
In order to use MSFV (without oversampling), we first adapt this method for OcTree meshes, as the original version is derived for tensor meshes only (cf. \cite{Haber2014a}). In this case, the total average run time per single simulation is 74 s on the same machine. The real and imaginary parts of $\Delta B^z$ at 100 Hz are shown in Figure~\ref{fig:rMSFV} and Figure~\ref{fig:iMSFV}, respectively.

% Comparison with MFV method on coarse mesh using homogenized (averaged) conductivity models 
We also carry out MFV simulations using homogenized electrical conductivity models that we construct using arithmetic, geometric and harmonic averaging of the fine-mesh conductivity inside each coarse cell of the OcTree mesh shown in Figure~\ref{fig:coarsedModel}.  
% Purpose of comparison
Doing so allow us to compare the accuracy that MFV solutions achieve on the coarse mesh with the one achieved by MSFV with and without oversampling.
The total average run time per single simulation is 124 s on the same machine.  The real and imaginary parts of $\Delta B^z$ at 100 Hz for each of the three homogenized solution are shown in Figures~\ref{fig:realResults}(d,e,f) and Figures~\ref{fig:imagResults}(d,e,f), respectively.  

% Norm of reference data as frequency varies

% Error computation and trend observations from table
Table \ref{tab:results} shows the relative errors in $l_2$ norm for the total, real and imaginary parts of $\Delta B^z$ obtained from comparing the reference (fine-mesh) solution with the MSFV, MSFV+O and MFV with three different homogenized solutions for each frequency and local extended domain size.  From this table, we observe the following.
% 0: Why we have large errors and why is it OK to accept them? (I mean the smaller error is in the order of 60%)
% 1: Oversampling significantly improves accuracy in the three parts of the solution considered.  In particular, MFV(geometric) better than MFV
First, our oversampling technique significantly improves the accuracy for the total response as well as for both the real and imaginary parts of the solution in comparison to the MSFV and MFV with three different homogenized conductivity models as the errors decrease with oversampling.  In particular, it is surprising to see how well MFV with simple geometric averaging did when compared with MSFV.
% 2: Increasing size of local oversampling area improves accuracy but is more expensive, yet still cheaper than ref solution.  
Second, as the size of the local extended domain increases the error decreases at the expense of more computational run time, which, however, is still considerably lower compared to the time of the reference solution for the cases of 2 and 4 padding cells. These results suggest that by using a local extended domain size of at least half the number of fine cells contained in the coarse cell(s) where the major contrast of conductivity happens, we may increase significantly the accuracy obtained with MSFV+O.
% 3: Discuss error behavior for real and imaginary parts as frequency varies
% Can we enhance this point with a more physical argument?  
Third, the errors for the imaginary part of $\Delta \B$ for the frequencies of 1, 10, 20, and 40 Hz resemble the errors for total $\Delta \B$ due to the real part of $\Delta \B$ is up to two orders of magnitude smaller than their corresponding imaginary parts (Figure \ref{fig:normRefSol}).  For the frequencies of 100 and 400 Hz, where the real and imaginary parts of $\Delta \B$ are roughly in the same order of magnitude, we see the relative error for total $\Delta \B$ represents the error contributions of these two parts of the data. 
% 4: Explain error behavior for 200 Hz:  high error at imaginary part, and zigzag in error
% Why for 200 Hz we obtained mostly a real signal and for the 400 Hz we don't see that?  
Fourth, for the simulation at 200 Hz the errors for the real part of $\Delta \B$ resemble the errors for total $\Delta \B$ due to the imaginary part of $\Delta \B$ is one order of magnitude smaller (Figure \ref{fig:normRefSol}).  The large errors in the imaginary part of the data as well as the increase and decrease in the error going from padding cell 2 to 4 and 4 to 8, respectively, may be related to the combined discretization error in simulating secondary field data and the extreme coarsening in the mesh. For this frequency, the cell sizes used to discretize the mineral deposit are in the order of the skin depth.
% 5: Explain error behavior of real part for 400 Hz
Fifth, for a simulation at 400 Hz with an extended domain size of 8 padding cells the slight increment in the error of the real part may be caused by excessive coarsening in the mesh for such high frequency.  In this case, the coarse cells used to discretize the mineral deposit are larger than the skin depth. Despite the excessive coarsening, our oversampled approximation yields comparable results to the reference (fine-mesh) solution for the largest two frequencies.

% Table with figures real part

% Table with figures imag part

% Table of relative errors
\begin{table*}[!htb]
\centering
\caption{Relative errors in $l_2$ norm for the total, real and imaginary parts of $\Delta \B$. \label{tab:results}}
\begin{subtable}{\textwidth}
\centering
\begin{tabular}{  l c c c c c c c}
  \toprule
  \multicolumn{8}{c}{Table of relative errors in $l_2$ norm} \\ \cmidrule(r){1-8}
  \multicolumn{1}{c}{} &
    \multicolumn{7}{c}{Frequency} \\
    \multicolumn{1}{c}{Method} & 
    \multicolumn{1}{c}{1 Hz} &
    \multicolumn{1}{c}{10 Hz} &
    \multicolumn{1}{c}{20 Hz} &
    \multicolumn{1}{c}{40 Hz} &
    \multicolumn{1}{c}{100 Hz} &
    \multicolumn{1}{c}{200 Hz} &
    \multicolumn{1}{c}{400 Hz} \\
    \midrule
    \multicolumn{1}{c}{} & \multicolumn{7}{c}{Relative errors for total $\Delta \B$ (\%)} \\ 
  MFV + Arithmetic    &    192.05   &   191.60 &   190.43 &  187.70 &   183.25 &   171.94 &   149.34  \\
  MFV + Geometric     &     68.57   &    68.51 &    68.34 &   67.72 &    64.96 &    60.61 &    55.62  \\ 
  MFV + Harmonic    &     97.20   &    97.20 &    97.18 &   97.11 &    96.81 &    96.30 &    95.57  \\ 
    MSFV        &     69.70   &    69.72 &    69.79 &   70.32 &    73.09 &    72.65 &    63.93  \\ 
    MSFV+O (2 padding cells)   &     15.84   &    15.83 &    15.79 &   15.75 &    16.17 &    15.98 &    13.46  \\
  MSFV+O (4 padding cells)   &     13.31   &    13.33 &    13.38 &   13.60 &    14.48 &    14.63 &    11.36  \\
  MSFV+O (8 padding cells)   &     10.63   &    10.65 &    10.72 &   10.98 &    12.20 &    12.67 &    10.08  \\
  \multicolumn{1}{c}{} & \multicolumn{7}{c}{Relative errors for real part of $\Delta \B$ (\%)} \\ 
  MFV + Arithmetic    &    216.47   & 214.97   &   211.12 &  202.06 &    188.63 &  170.99 &   159.57 \\
  MFV + Geometric     &     81.01   &  80.95   &    80.75 &   79.72 &     73.05 &   60.33 &    34.18 \\ 
  MFV + Harmonic    &     98.46   &  98.45   &    98.43 &   98.33 &     97.74 &   96.52 &    93.45 \\ 
    MSFV        &     73.09   &  72.92   &    72.56 &   72.29 &     74.22 &   70.82 &    59.47 \\ 
    MSFV+O (2 padding cells)   &     21.41   &  21.34   &    21.16 &   20.57 &     18.11 &   14.21 &     8.25 \\
  MSFV+O (4 padding cells)   &     18.36   &  18.38   &    18.39 &   18.19 &     15.92 &   12.49 &     8.14 \\
  MSFV+O (8 padding cells)   &     16.12   &  16.10   &    16.02 &   15.49 &     12.74 &   10.40 &     8.51 \\  % 400Hz large error why?
  \multicolumn{1}{c}{} & \multicolumn{7}{c}{Relative errors for imaginary part of $\Delta \B$ (\%)} \\ 
  MFV + Arithmetic    &   192.05    &   191.26 &   189.26 &  184.63 &    174.80 &  197.75 &   133.27 \\
  MFV + Geometric     &    68.57    &    68.32 &    67.61 &   64.99 &     50.30 &   68.40 &    76.55 \\ 
  MFV + Harmonic    &    97.20    &    97.18 &    97.11 &   96.86 &     95.38 &   89.47 &    98.52 \\ 
    MSFV        &    69.70    &    69.67 &    69.64 &   69.91 &     71.34 &  114.01 &    69.83 \\ 
    MSFV+O (2 padding cells)   &    15.84    &    15.74 &    15.45 &   14.57 &     12.68 &   42.77 &    18.53 \\
  MSFV+O (4 padding cells)   &    13.31    &    13.24 &    13.06 &   12.46 &     11.99 &   43.84 &    14.81 \\  % 200 Hz large error why?
  MSFV+O (8 padding cells)   &    10.63    &    10.55 &    10.36 &    9.81 &     11.35 &   41.16 &    11.98 \\
  \bottomrule
 \end{tabular}
\end{subtable}
\end{table*}

\section{Conclusions}\label{sec:conclusions}
We develop an oversampling technique for the multiscale finite volume method (\cite{Haber2014a}) to simulate quasi-static electromagnetic responses in the frequency domain for geophysical settings that include highly heterogeneous media and features varying at different spatial scales. Simulating these types of geophysical settings requires both large CPU time and memory usage; they often need a very large and fine mesh to be discretized accurately.
Our method begins by assuming a coarse mesh nested into a fine mesh, which accurately discretizes the geophysical setting. For each coarse cell, we independently solve a local version of the original Maxwell's system subject to linear boundary conditions on an extended domain.  To solve the local Maxwell's system, we use the fine mesh contained in the extended domain and the Mimetic Finite Volume method.  Afterwards, these local solutions, called basis functions, together with a weak continuity condition are used to construct a coarse-scale version of the global problem that is much cheaper to solve.  
The proposed method produces results comparable to those obtained by simulating electromagnetic responses on a fine mesh using classical discretization methods, such as mimetic finite volume, while drastically reducing the size of the linear system of equations and the computational time. Using the oversampling technique in the presented example, the size of the coarse-mesh system is only about 10\% of the fine-mesh system size, while the relative er\-ror is less than 25\% for all the cases considered.

%------------------------------------------------------------
% Acknowledgments
% -----------------------------------------------------------
\begin{acknowledgements}
The authors would like to thank Dominique Fournier for providing the data that we used to construct the synthetic conductivity model used for the 3D numerical experiments presented, and Gudni Rosenkjar for his assistance with the software Paraview that we use to visualize the 3D numerical results. 
%We also thank the anonymous referees for their thoughtful and constructive comments that help to improve this paper.
The funding for this work is provided through the University of British Columbia's Four-Year-Fellowship program and NSERC Industrial Research Chair program.
\end{acknowledgements}
%------------------------------------------------------------
% Bibliography
% -----------------------------------------------------------
\clearpage
\bibliographystyle{spmpsci}      % mathematics and physical sciences
\bibliography{noURLlibrary}

\begin{thebibliography}{10}
\providecommand{\url}[1]{{#1}}
\providecommand{\urlprefix}{URL }
\expandafter\ifx\csname urlstyle\endcsname\relax
  \providecommand{\doi}[1]{DOI~\discretionary{}{}{}#1}\else
  \providecommand{\doi}{DOI~\discretionary{}{}{}\begingroup
  \urlstyle{rm}\Url}\fi

\bibitem{Alcouffe1981}
Alcouffe, R., Brand, A., Dendy, J., Painter, J.: {The Multi-grid method for the
  difussion equation with strongly discontinuous coefficients}.
\newblock Siam J. Sci. Stat. Comput. \textbf{2}(4), 430--454 (1981).
\newblock \doi{http://epubs.siam.org/doi/abs/10.1137/0902035}

\bibitem{Amestoy2001}
Amestoy, P., Duff, I., L'Excellent, J.Y., Koster, J.: {MUMPS: a general purpose
  distributed memory sparse solver}.
\newblock Int. Work. Appl. Parallel Comput. Springer Berlin Heidelb. pp.
  121--130 (2001)

\bibitem{Benner2015}
Benner, P., Willcox, K.: {A Survey of Projection-Based Model Reduction Methods
  for Parametric Dynamical Systems ∗}.
\newblock SIAM Rev. \textbf{57}(4), 483--531 (2015)

\bibitem{Dendy1982}
Dendy, J.: {Black box multigrid}.
\newblock J. Comput. Phys. \textbf{48}(3), 366--386 (1982).
\newblock \doi{10.1016/0021-9991(82)90057-2}

\bibitem{Efendiev2007}
Efendiev, Y., Hou, T.Y.: {Multiscale finite element methods for porous media
  flows and their applications}.
\newblock Appl. Numer. Math. \textbf{57}(5-7), 577--596 (2007).
\newblock \doi{10.1016/j.apnum.2006.07.009}

\bibitem{Efendiev2009}
Efendiev, Y., Hou, T.Y.: {Multiscale finite element methods: theory and
  applications}.
\newblock Springer New York (2009).
\newblock \doi{10.1007/978-0-387-09496-0}

\bibitem{Haber2014b}
Haber, E.: {Advanced Modeling and Inversion Tools for Controlled Source EM}.
\newblock In: 76th EAGE Conf. Exhib., June 2014, pp. WS9--B02. EAGE (2014).
\newblock \doi{10.3997/2214-4609.20140555}

\bibitem{Haber2014}
Haber, E.: {Computational Methods in Geophysical Electromagnetics}.
\newblock Society for Industrial and Applied Mathematics (2014)

\bibitem{Haber2001}
Haber, E., Ascher, U.: {Fast finite volume simulation of 3D electromagnetic
  problems with highly discontinuous coefficients}.
\newblock SIAM J. Sci. Comput. \textbf{22}(6), 1943--1961 (2001).
\newblock \doi{10.1137/S1064827599360741}

\bibitem{Haber2007a}
Haber, E., Heldmann, S.: {An octree multigrid method for quasi-static Maxwell's
  equations with highly discontinuous coefficients}.
\newblock J. Comput. Phys. \textbf{223}(2), 783--796 (2007).
\newblock \doi{10.1016/j.jcp.2006.10.012}

\bibitem{Haber2014a}
Haber, E., Ruthotto, L.: {A multiscale finite volume method for Maxwell's
  equations at low frequencies}.
\newblock Geophys. J. Int. \textbf{m}, 1--20 (2014)

\bibitem{Hajibeygi2009}
Hajibeygi, H., Jenny, P.: {Multiscale finite-volume method for parabolic
  problems arising from compressible multiphase flow in porous media}.
\newblock J. Comput. Phys. \textbf{228}, 5129--5147 (2009).
\newblock \doi{10.1016/j.jcp.2009.04.017}

\bibitem{Hiptmair1998}
Hiptmair, R.: {Multigrid Method for Maxwell's Equations}.
\newblock SIAM J. Numer. Anal. \textbf{36}(1), 204--225 (1998).
\newblock \doi{10.1137/S0036142997326203}

\bibitem{Horesh2011}
Horesh, L., Haber, E.: {A Second Order Discretization of Maxwell's Equations in
  the Quasi-Static Regime on OcTree Grids}.
\newblock SIAM J. Sci. Comput. \textbf{33}(5), 2805--2819 (2011).
\newblock \doi{10.1137/100798508}

\bibitem{Hou2003}
Hou, T.Y.: {Numerical Approximations to Multiscale Solutions in Partial
  Differential Equations}.
\newblock In: Front. Numer. Anal., pp. 241--301. Springer (2003)

\bibitem{Hou1997}
Hou, T.Y., Wu, X.: {A Multiscale Finite Element Method for Elliptic Problems in
  Composite Materials and Porous Media}.
\newblock J. Comput. Phys. \textbf{134}(1), 169--189 (1997).
\newblock \doi{10.1006/jcph.1997.5682}

\bibitem{Hou1999}
Hou, T.Y., Wu, X., Cai, Z.: {Convergence of a Multiscale Finite Element Method
  for Elliptic Problems with Rapidly Oscillating Coefficients}.
\newblock Math. Comput. Am. Math. Soc. \textbf{68}(227), 913--943 (1999).
\newblock \doi{10.1155/2013/764165}

\bibitem{Hyman1998}
Hyman, J.M., Shashkov, M.: {Mimetic discretizations for Maxwell equations and
  the equations of magnetic diffusion}.
\newblock Tech. rep., United States. Department of Energy. Office of Energy
  Research (1998)

\bibitem{Hyman1999}
Hyman, J.M., Shashkov, M.: {Mimetic discretizations for Maxwell's equations}.
\newblock J. Comput. Phys. \textbf{151}(2), 881--909 (1999).
\newblock \doi{10.1006/jcph.1999.6225}

\bibitem{Hyman1999a}
Hyman, J.M., Shashkov, M.: {The orthogonal decomposition theorems for mimetic
  finite difference methods}.
\newblock SIAM J. Numer. Anal. \textbf{36}(3), 788--818 (1999).
\newblock \doi{10.1137/S0036142996314044}

\bibitem{Jacobsson2007}
Jacobsson, P.: {Nedelec elements for computational electromagnetics} (2007)

\bibitem{Jenny2003}
Jenny, P., Lee, S., Tchelepi, H.: {Multi-scale finite-volume method for
  elliptic problems in subsurface flow simulation}.
\newblock J. Comput. Phys. \textbf{187}(1), 47--67 (2003).
\newblock \doi{10.1016/S0021-9991(03)00075-5}

\bibitem{Jin2002}
Jin, J.: {The Finite Element Method in Electromagnetics}, 2nd edn.
\newblock Wiley, New York (2002)

\bibitem{Key2011a}
Key, K., Ovall, J.: {A parallel goal-oriented adaptive finite element method
  for 2.5D electromagnetic modelling}.
\newblock Geophys. J. Int. \textbf{186}(1), 137--154 (2011).
\newblock \doi{10.1111/j.1365-246X.2011.05025.x}

\bibitem{Lipnikov2004}
Lipnikov, K., Morel, J., Shashkov, M.: {Mimetic finite difference methods for
  diffusion equations on non-orthogonal non-conformal meshes}.
\newblock J. Comput. Phys. \textbf{199}, 589--597 (2004).
\newblock \doi{10.1016/j.jcp.2004.02.016}

\bibitem{MacLachlan2006}
MacLachlan, S., Moulton, J.D.: {Multilevel upscaling through variational
  coarsening}.
\newblock Water Resour. Res. \textbf{42}(2), W02,418:1--9 (2006).
\newblock \doi{10.1029/2005WR003940.1.}

\bibitem{MacLachlan2012}
MacLachlan, S., Moulton, J.D., Chartier, T.P.: {Robust and adaptive multigrid
  methods: comparing structured and algebraic approaches}.
\newblock Numer. Linear Algebr. with Appl. \textbf{19}(2), 389--413 (2012).
\newblock \doi{10.1002/nla}

\bibitem{Monk2003}
Monk, P.: {Finite element methods for Maxwell's equations}.
\newblock Clarendon (2003).
\newblock \doi{10.1093/acprof:oso/9780198508885.001.0001}

\bibitem{Oldenburg1990}
Oldenburg, D.W.: {Inversion of electromagnetic data: an overview of new
  techniques}.
\newblock Surv. Geophys. \textbf{11}(2-3), 231--270 (1990).
\newblock \doi{10.1007/BF01901661}

\bibitem{Pavliotis2008}
Pavliotis, G., Stuart, A.: {Multiscale methods: averaging and homogenization}.
\newblock Springer New York (2008).
\newblock \doi{10.1007/978-0-387-73829-1}

\bibitem{Schwarzbach2009}
Schwarzbach, C.: {Stability of finite element discretization of Maxwell's
  equations for geophysical applications}.
\newblock Ph.D. thesis, University of Frieberg, Germany (2009)

\bibitem{Ward1988}
Ward, S.H., Hohmann, G.W.: {Electromagnetic theory for geophysical
  applications}.
\newblock In: M.N. Nabighian (ed.) Electromagn. methods Appl. Geophys., vol.~1,
  chap.~4, pp. 130--311. Society of Exploration Geophysicists (1988)

\bibitem{Yang2014}
Yang, D., Fournier, D., Oldenburg, D.W.: {3D Inversion of EM data at Lalor
  mine: in pursuit of a unified electrical conductivity model}.
\newblock In: Explor. Deep VMS Ore Bodies Hudbay Lalor Case Study, pp. 1--4. BC
  Geophysical Society (2014)

\bibitem{Yee1966}
Yee, K.: {Numerical Solution of Initial Boundary Value Problems Involving
  Maxwell's Equations in Isotropic Media}.
\newblock Antennas Propagation, IEEE Trans. \textbf{14}(3), 302--307 (1966)

\bibitem{Zhdanov2010}
Zhdanov, M.S.: {Electromagnetic geophysics: Notes from the past and the road
  ahead}.
\newblock Geophysics \textbf{75}(5), 75A49--75A66 (2010).
\newblock \doi{10.1190/1.3483901}

\end{thebibliography}

% Trick to remove url or other field in .bib
% grep -v "url =" library.bib > noURLlibrary.bib

\end{document}